\documentclass[usenatbib,twocolumn]{mn2e}
\usepackage{graphicx}
\usepackage{amsfonts}
\usepackage{epsfig,rotating,natbib,pstricks}
\usepackage{color}
\usepackage{amssymb,latexsym}
\usepackage{amsmath,wasysym}
\usepackage{verbatim}
\usepackage[T1]{fontenc}
\usepackage{aecompl}

\def\beq{\begin{equation}}
\def\eeq{\end{equation}}
\def\baq{\begin{eqnarray}}
\def\eaq{\end{eqnarray}}

\def\p3m{P$^3$M}
\def\ap3m{AP$^3$M}

\def\h1{H\/I}
\def\omegah1{\Omega_{\h1}}
\def\ph1{P_{_{\h1}}}
\def\ph1k{P_{_{\h1}}(k)}
\def\dh1k{\Delta^2_{_{\h1}}(k)}

\def\msun{M_{\odot}}

\def\mblack2{{MassiveBlack-II }}
\def\mblack{{MassiveBlack }}

\def\mhi{M_{\text{HI}}}
\def\dmax{D_{\text{max}}}
\def\vmax{V_{\text{max}}}
\def\w50{W_{50}}
\def\s21{S_{21}}

\newcommand{\be}{\begin{equation}}
\newcommand{\e}{\end{equation}}

\title[The HI Mass Function]{The Population of Galaxies that Contribute to The 
  HI Mass Function}

\author[]{\parbox{18cm}{Saili Dutta$^{1}$\thanks{E-mail: sailidutta@niser.ac.in(SD); nkhandai@niser.ac.in(NK); biprateep@pitt.edu(BD)},Nishikanta Khandai$^{1\star}$,Biprateep Dey$^{1,2\star}$}
  \vspace{0.3cm}\\
  $^{1}$ {School of Physical Sciences, National Institute of Science Education and Research, HBNI, Jatni 752050, India}\\
  $^{2}$ {Pittsburgh Particle Physics, Astrophysics and Cosmology Center (PITT PACC) and Department of Physics and Astronomy,}\\
       \quad \quad {University of Pittsburgh, 3941 O'Hara Street, Pittsburgh, PA 15260, USA}
}
\label{firstpage}

\def\LaTeX{L\kern-.36em\raise.3ex\hbox{a}\kern-.15em
    T\kern-.1667em\lower.7ex\hbox{E}\kern-.125emX}

\pagerange{\pageref{firstpage}--\pageref{lastpage}}

\begin{document}

\maketitle

\begin{abstract}
  We look at the contribution of different galaxy populations 
  to the atomic hydrogen (HI) mass function (HIMF) and the HI density parameter,
   $\Omega_{\text{HI}}$, in the local Universe. Our analysis is based on 
  a sample of 7857 HI-selected galaxies selected from a volume common to the SDSS and 
  ALFALFA (40\% catalog -- $\alpha.40$) surveys. We define different populations 
  of galaxies in the color(\emph{u-r})-magnitude($M_{\text{r}}$) 
  plane and compute the HIMF for each 
  of them. Additionally we compute the HIMF for dark galaxies; these are 
  undetected in SDSS and  represent $\sim 2\%$ of the total sample. We find that the 
  luminous red population dominates the total HIMF for 
  $\log_{10}(M_{\text{HI}}h^2_{70}/\msun) \geq 10.4$. The full red population 
  -- luminous and faint -- represents about  $\sim 17\%$ of the 
  $\Omega_{\text{HI}}$ budget,  while that of the dark population
  is $\sim 3\%$. The HIMF about the knee, 
  $\log_{10}(M_{\text{HI}}h^2_{70}/\msun) \in [8,10.4]$, is dominated by the 
  faint and luminous blue populations, the latter dominating at larger masses
  in this interval.  Their total contribution to $\Omega_{\text{HI}}$ is 
  $\sim 55-70\%$, the variation depending on the definition 
  of population. The dominant populations at the low mass end, 
  $\log_{10}(M_{\text{HI}}h^2_{70}/\msun) \leq 8.0$ are the faint blue 
  and faint bluer populations, the latter's dominance being sensitive to its definition.
  The full blue (blue--bluer luminous and faint) population 
  represents $\sim 80\%$ of $\Omega_{\text{HI}}$. A bimodal HIMF suggested by our results
  is however not seen since the amplitude of the HIMF of the luminous red population 
  is small compared to that of the luminous blue population.

\end{abstract}



\section{Introduction}
The relationship between gas, metals, feedback and stars in galaxies 
is crucial for our understanding of galaxy formation and evolution. 
We need a clearer picture of how dark matter halos, which harbor 
galaxies, are supplied with cold gas, the fuel for star formation;
how local and global conditions in the galaxy are responsible 
in processing them into stars; and finally how the gas is polluted with metals and 
recycled back to the intergalactic medium due to feedback processes
within the galaxy. 

Three basic properties that describe galaxies are: (i) the star formation rate (SFR), (ii)
the stellar mass ($M_{\text{star}}$) and (iii) the cold neutral hydrogen gas mass both in molecular ($M_{\text{H}_2}$) and atomic ($M_{\text{HI}}$) phases. The amount of neutral gas tells
us the amount of fuel currently available for future star formation. 
The SFR 
is the current rate of forming stars from the supply of cold gas that is available,
while $M_{\text{star}}$ depends on the integrated star formation history. 
Within each galaxy, at sub-kpc scales, the observed correlation between 
surface density of molecular hydrogen, 
$\Sigma_{\text{H}_2}$, and the  SFR surface density, 
$\Sigma_{\text{SFR}}$, is stronger \citep{2008AJ....136.2846B,2008AJ....136.2782L} 
as compared to the correlation between  the HI surface density, 
$\Sigma_{\text{HI}}$, and $\Sigma_{\text{SFR}}$. The HI is often distributed 
beyond the optical radius of the galaxy and is more diffuse 
whereas star formation and $\text{H}_2$ are mostly concentrated within the optical 
radius and occur in clumpier regions \citep{2008AJ....136.2782L}. However when the HI gas
cools and becomes denser it transitions to molecular gas, which cools further
and becomes denser leading eventually 
to gravitational collapse to form stars. It is therefore common to 
correlate the total gas (HI+$\text{H}_2$) to the SFR, the so-called Kennicutt-Schmidt law
\citep{1959ApJ...129..243S,1963ApJ...137..758S,1998ApJ...498..541K,1989ApJ...344..685K} 
for star formation, where $\Sigma_{\text{SFR}} \propto  \Sigma_{\text{gas}}^{1.4}$.

With the aid of multiwavelength observations in the 
optical, ultraviolet (UV) and infrared (IR) bands 
followed up by spectroscopic measurements, we are able to infer scaling relations 
between various properties of galaxies, derived and/or observed. These scaling relations 
are very important since any theoretical model of galaxy formation should be able to 
reproduce them. In practice, these relations are used as parameters in theoretical models
of galaxy formation when studied in cosmological volumes, 
since the physics on small, subparsec scales is never resolved 
\citep{2003MNRAS.339..289S,2005Natur.433..604D}. 
A multi-pronged approach is used to study the distribution of HI in the 
post reionization Universe; these include cosmological hydrodynamical simulations
of galaxy formation \citep{2017MNRAS.467..115D}, semi-analytical models 
(SAM) of galaxy formation \citep{2017MNRAS.465..111K} and halo-occupation distribution 
(HOD) \citep{2018MNRAS.479.1627P}. All of these approaches invoke in some form or other
some of the observed scaling relations and model predictions 
are tested against observations which were not used as their input.
Since these scaling relations depend on the observed sample they may be biased, i.e.
the relations may depend on how the sample is chosen. E.g.  
the $M_{\text{HI}}-M_{\text{star}}$ relation differs if a sample is chosen by HI mass
\citep{2012ApJ...756..113H}
or stellar mass \citep{2010MNRAS.403..683C}. On the other hand quantities like the 
luminosity function, mass function, correlation function (to name a few) 
are corrected for the survey selection and tell us about the underlying abundance and 
distribution of different galaxy types in the survey volume. 

By analyzing data from optical, UV and IR surveys over the past decade we have formed 
a clearer picture of how galaxies, on average, have formed and 
evolved over the past $\sim$ 12.5 billion years from 
redshift $z=6$ to today. The cosmic stellar mass density, $\rho_*$ (units of 
$\msun \text{Mpc}^{-3}3$), has increased monotonically by nearly 2.5 decades 
from $\log(\rho_*) \simeq 6.3$ at $z=6$
to  $\log(\rho_*) \simeq 8.8$ today \citep{2014ARA&A..52..415M}. 
These observations also tell us how the cosmic SFR density 
(SFRD denoted by $\psi$ with units $\msun \text{yr}^{-1} \text{Mpc}^{-3}$) has changed 
during this time. It increases from $\log(\psi) = -1.7$ at $z=6$ to a peak value of 
$\log(\psi) = -0.9$ at $z\simeq 2$ and finally dropping by a decade  
to $\log(\psi) = -1.8$ today 
\citep{2005ApJ...632..169L,2014ARA&A..52..415M,2014PhR...541...45C}.
In contrast, surveys targeting gas content of galaxies have lagged behind in depth 
and number. The 21cm line of HI being a weak line its detection in emission 
is limited only to the local Universe. 

Blind HI surveys like the 
HI Parkes All Sky Survey  \citep[HIPASS,][]{2004MNRAS.350.1195M}
and the Arecibo Fast Legacy ALFA survey \citep[ALFALFA,][]{2005AJ....130.2598G}
have been used to accurately measure the HIMF 
\citep{2003AJ....125.2842Z,2005MNRAS.359L..30Z,2010ApJ...723.1359M,
  2011AJ....142..170H,2018MNRAS.477....2J} in the local Universe ($z \leq 0.05$).
The HIMF can then be integrated to obtain the HI density parameter $\Omega_{\text{HI}}$.
At higher redshifts the 21cm flux gets further diluted  and 
direct detection becomes difficult with existing instruments. Stacking the HI data
on known optical counterparts or alternately cross-correlating the HI intensity maps 
with the optical catalog in a common volume,
then becomes a useful tool in making detections. The stacking method 
has been applied for star-forming galaxies at $z=0.24$ \citep{2007MNRAS.376.1357L},  
galaxies in a cluster environment at $z = 0.37$ 
\citep{2009MNRAS.399.1447L}, for field galaxies at $z=0.1-0.2$ \citep{2013MNRAS.435.2693R},
in the zCOSMOS\footnote{http://cesam.lam.fr/zCosmos/} 
field at $z=0.37$ \citep{2016MNRAS.460.2675R}
and the VVDS\footnote{https://cesam.lam.fr/vvds/} 
field at $z=0.32$ \citep{2018MNRAS.473.1879R}. 
Most of the HI observations using the stacking method were done with the 
Giant Metrewave Radio Telescope (GMRT).
More recently
stacking has been applied to estimate the HI content 
in filaments \citep{2019MNRAS.489..385T}. However at higher redshifts ($z\sim 1.3$)
the stacking method did not result in a detection of star forming galaxies 
in the DEEP2\footnote{http://deep.ps.uci.edu/} 
field \citep{2016ApJ...818L..28K} whereas positive detections have 
been reported by cross-correlating 
the HI Intensity map with an optical survey at $z\sim 0.8$ 
\citep{2010Natur.466..463C,2013ApJ...763L..20M}. In the stacking method 
one can estimate $\Omega_{\text{HI}}$ after correcting for the optical survey's completeness 
limit, whereas the cross-correlation method constrains $\Omega_{\text{HI}}b_{\text{HI}}r$.
Here $b_{\text{HI}}$ and $r$ are the HI bias parameter and galaxy-HI cross 
correlation coefficient. In summary these HI surveys  constrain $\Omega_{\text{HI}} = 4 \pm 1.6 \times 10^{-4}$ out to $z \simeq 0.4$ \citep{2018MNRAS.473.1879R}.

Beyond $z = 0.4$ and out to $z \simeq 5$ the HI content of the Universe is derived from studying damped 
Lyman-$\alpha$ systems (DLA) seen, in absorption, in the spectra of quasars
\citep{2005ApJ...635..123P,2012A&A...547L...1N,2016ApJ...818..113N,2017MNRAS.471.3428R}.
Combining both these approaches at low redshift (emission) and high redshift (absorption)
one sees that $\Omega_{\text{HI}}$ increases monotonically as $(1+z)^{0.56}$ from 
 $\Omega_{\text{HI}}(z=0) = 4\times 10^{-4}$ to  $\Omega_{\text{HI}}(z=5) = 1.1\times 10^{-3}$
\citep[for a full compilation of all observational results see][]{2018MNRAS.473.1879R}.
The HI density at $z=0$ is only $1.5\times$ smaller than at $z=2$ with the data 
consistent with a no-evolution picture. On the other hand the cosmic SFRD has decreased 
10 fold in this interval. Clearly our picture of galaxy formation is incomplete
at these redshifts since the decrease in SFRD is not 
commensurate with a depletion of HI. Upcoming HI surveys like the 
Square Kilometre Array (SKA) will help us in this direction.

However even with the existing data it is important to understand 
the dependence of the abundance and distribution of HI selected galaxies 
on different galaxy properties and environments.
The clustering of HI selected galaxies in the ALFALFA survey 
has been measured \citep{2012ApJ...750...38M,2013ApJ...776...43P,2017ApJ...846...61G}
suggesting that ALFALFA galaxies cluster weakly. The dependence of HIMF has also 
been explored on the environment with \cite{2018MNRAS.477....2J} reporting 
a decrease in the low mass slope with increasing density in ALFALFA 
whereas \cite{2019MNRAS.486.1796S} find that the low mass slope increases with increasing 
density in the HI Zone of Avoidance survey with the Parkes telescope. Similarly the 
the HI velocity width function has been studied for wall and void galaxies as well as red 
and blue galaxies \citep{2014MNRAS.444.3559M}, indicating a strong dependence 
on both environment and galaxy color. \citet{2003AJ....125.2842Z} have 
looked at the dependence of the HIMF on galaxy morphology as well  
as on the early-late type classification. In this paper we look at the dependence 
of the HIMF on the different populations classified from the color-magnitude plane.
This is similar in spirit to the analysis of  
\citet{2003AJ....125.2842Z,2014MNRAS.444.3559M} carried out in HI surveys 
and also bears resemblance to similar analysis carried out for optical surveys
where the contribution of the red and blue galaxies to the galaxy stellar mass function
and luminosity functions are explored \citep{2004ApJ...600..681B,2009ApJ...707.1595D}.

Our paper is organized as follows. In section~\ref{sec_data} we describe a sample common 
to ALFALFA and SDSS (Sloan Digital Sky Survey), in section~\ref{sec_results} 
we report our measurements of the HIMF 
for different populations within this sample, in section~\ref{sec_discussion} 
we discuss our results and their implication and finally we summarize and conclude in 
section~\ref{sec_summary}. We assume a flat $\Lambda $CDM cosmology with $\Omega_m = 0.3$,
$\Omega_\Lambda = 0.7$ and a value for the dimensionless Hubble constant  
$h_{70} = 0.7$.

\section{Data}
\label{sec_data}

At 40\% data release \citep{2011AJ....142..170H} 
ALFALFA has surveyed over $\sim$2752 deg$^{2}$ of the sky,
with comoving volume of $\sim$2.65 $\times 10^6$ Mpc$^3$.
Covering 40\% of the targeted survey area, 
the $\alpha.40$ catalog contains 15855 sources in the regions 
$7^h 30^m<$R.A.$<16^h 30^m$, 
$4^{\circ}<$ dec.$<16^{\circ}$, and 
$24^{\circ}<$ dec.$<28^{\circ}$ and
$22^h<$R.A.$<3^h$, $14^{\circ}<$ dec.$<16^{\circ}$, and
$24^{\circ}<$ dec.$<32^{\circ}$.
Most of these objects have optical counterparts in 
the Sloan Digital Sky Survey (SDSS) Data Release 7 (DR7) \citep{2009ApJS..182..543A}.

This catalog contains the following observed quantities  
(i) an unique entry number from the Arecibo General Catalog(AGC);
(ii) right ascension  and declination  of the HI source and the most 
probable optical counterpart (OC) from SDSS DR7;
(iii) heliocentric velocity ($cz_{helio}$) which is the
 midpoint of the HI flux density profile;
(iv) velocity profile width ($\w50$), measured as the full width at 50\% of the 
peak  HI flux density;
(v) the integrated flux density of the HI source ($\s21$). 
The derived quantities, contained in this catalog, are - 
(i) distance to the object in Mpc (D). For sources with $cz_{helio} > 6000 $ km s$^{-1}$ 
this quantity is $cz_{\text{\tiny{CMB}}}/H_0$, 
where $cz_{\text{\tiny{CMB}}}$ is the velocity in the cosmic microwave background (CMB) 
reference frame and $H_0$ is the Hubble constant (taken to be 70 km s$^{-1}$ Mpc$^{-1}$). 
For sources with $cz_{helio} < 6000 $ km s$^{-1}$ this quantity has been estimated 
using a local flow model \citep{2005PhDT.........2M}
(ii) the HI mass ($\mhi$), computed as 
\beq
\frac{\mhi}{\msun} = 2.356\times10^5 \left(\frac{D}{\text{Mpc}}\right)^2 
\left(\frac{\s21}{\text{Jy.km.s}^{-1}}\right)
\label{eq_massfluxrelation}
\eeq
The $\alpha.40$ catalog also provides another property which is the Code number, 
with a value of 1,2 or 9. The sources with SNR $>6.5$ are referred as Code 1 objects, 
Code 2 objects are the detections with SNR $<6.5$ 
and Code 9 refers to the high velocity clouds (HVC).

For our analysis we have considered only Code 1 galaxies.
The number of Code 1 galaxies is 11941.
We have also considered a cut for $cz_{\text{\tiny{CMB}}} < 15000 \text{km.s}^{-1}$
to avoid radio frequency interference (RFI) 
generated by Federal Aviation Administration (FAA) 
radar at the San Juan airport 
\citep{2010ApJ...723.1359M, 2011AJ....142..170H}. 
This restricts the sample to redshift, $z \leq 0.05$ 
and reduces the sample to 10785 galaxies.

\begin{figure}
  \begin{tabular}{c}
    \includegraphics[width=\columnwidth]{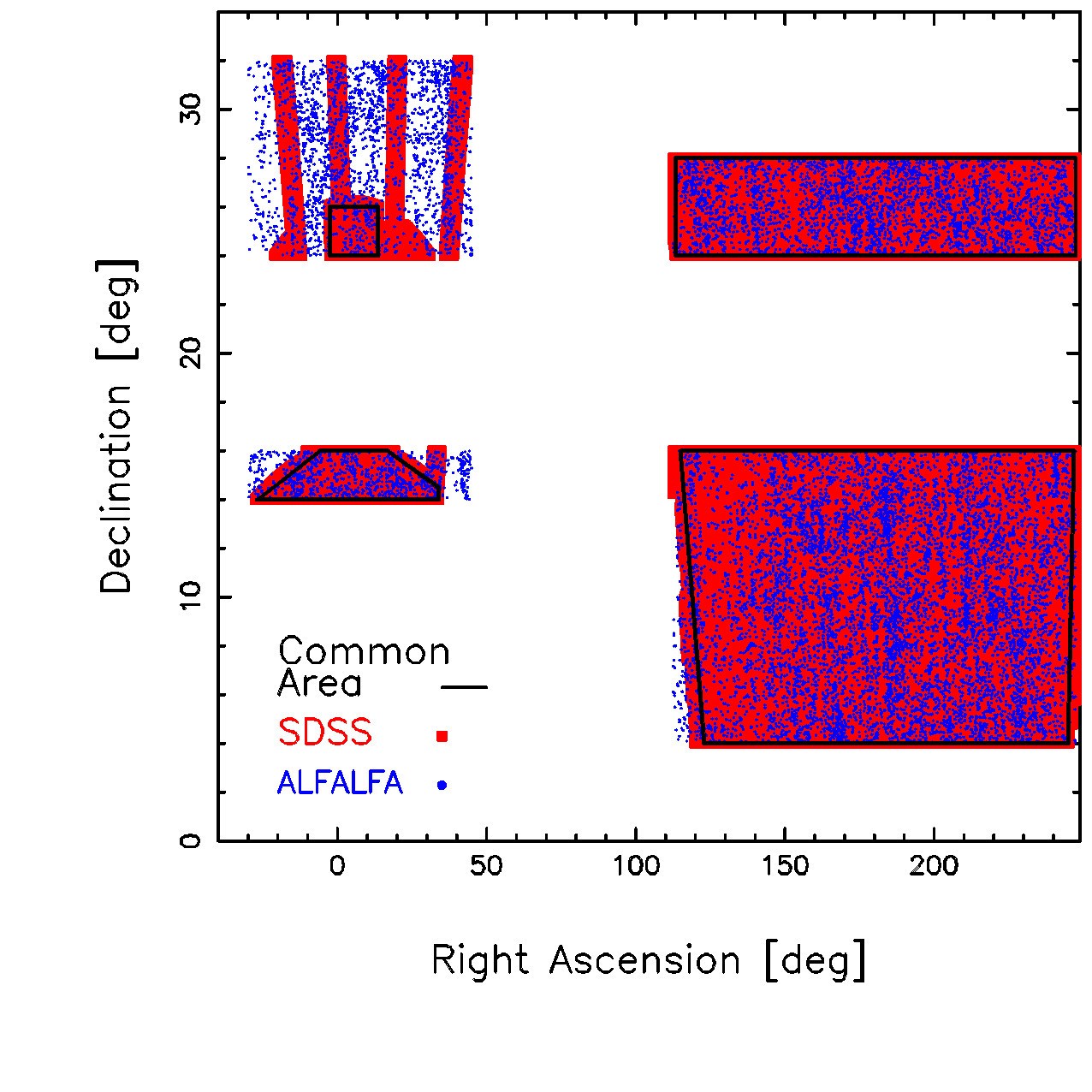} \\
  \end{tabular}
  \caption{Common footprint of SDSS and ALFALFA used in this work.
    The pale grey patches is that of SDSS DR7 overlapping with the ALFALFA survey region.
    The black dots are the positions of ALFALFA galaxies. The common boundary 
used in this work is outlined by the thick black line.}
  \label{fig_radec}
\end{figure}

Since ALFALFA is not totally overlapping with the SDSS footprint, 
we have defined a common boundary for both the surveys in this work in 
figure~\ref{fig_radec}. 
For this work our final area of analysis has four subregions whose vertices 
right ascension-declination (RA,dec) 
are given as 
(i)($123^{\circ},4^{\circ}$), ($245^{\circ},4^{\circ}$), 
($247^{\circ},16^{\circ}$), ($115^{\circ},16^{\circ}$);
(ii) ($113.31^{\circ},24^{\circ}$), ($247.5^{\circ},24^{\circ}$), 
($247.5^{\circ},28^{\circ}$), ($113.31^{\circ},28^{\circ}$);
(iii) ($-27^{\circ},14^{\circ}$), ($34^{\circ},14^{\circ}$), 
($34^{\circ},14.5^{\circ}$), ($17^{\circ},16^{\circ}$), ($-6^{\circ},16^{\circ}$);
and (iv) ($-2.6^{\circ},24^{\circ}$), ($13.6^{\circ},24^{\circ}$), 
($13.6^{\circ},26^{\circ}$), ($-2.6^{\circ},26^{\circ}$).
The total volume of these regions is  $2.02 \times 10^6$  Mpc$^3$ and it corresponds 
to an angular area of  $\sim 2093$ deg$^2$. In this region there are 8344 
galaxies in ALFALFA. This includes removal of 4 OH high-redshift 
impostors \citep{2016MNRAS.459..220S}.

\begin{figure}
\centering
\includegraphics[width=\columnwidth]{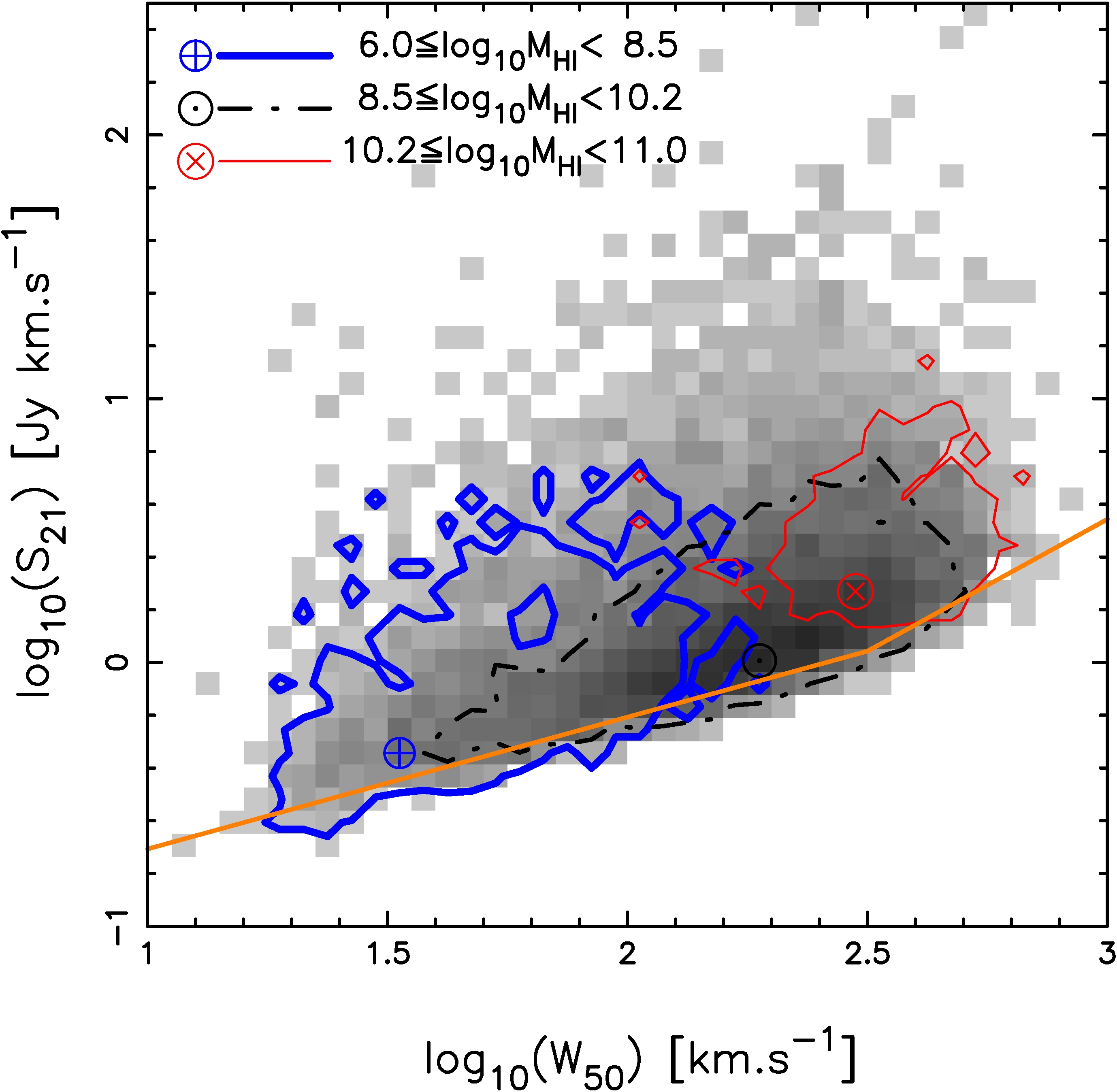}
\caption{The distribution of galaxies in the $\s21-\w50$ plane is shown for all Code 1  
ALFALFA galaxies. The broken solid line is the $50\%$ completeness 
relation \citep{2011AJ....142..170H}. It is given in 
eq.~\ref{eq_completeness}. We have further divided the sample into three mass bins: 
$\log_{10} [\mhi/\msun] \in [6.0,8.5[, [8.5,10.2[, [10.2,11.0[$. The $1\sigma$ contours 
and the peaks of the distributions for these three populations 
are given by thick solid line (plus-circle), the dot-dashed line (dot-circle) and 
the thin solid line (cross-circle).}
\label{fig_sw}
\end{figure}

In figure~\ref{fig_sw} we look at the distribution of Code 1 objects in ALFALFA
in the $\s21-\w50$ plane. We also show the distributions in three mass bins which 
correspond 
to the small-mass end or faint-end (thick solid line), 
the knee (dot-dashed line) and the high-mass end (thin-solid line).
As seen from these distributions on  
average the velocity width increases with increasing mass.
This is expected and is also seen in the $\mhi-\w50$ relation
\citep[see figure~7 of][]{2014MNRAS.444.3559M}. One also sees that it is very 
unlikely to have a low(high) mass galaxy with a large(small) velocity width. 
However intermediate mass objects can have the full range of velocity widths. 

The broken solid line in figure~\ref{fig_sw} is the sensitivity limit and 
is given by a $50\%$ completeness relation of Code 1 objects in 
eq.~\ref{eq_completeness} \citep{2011AJ....142..170H}.
This tells us that the detection of objects not only depends on 
the integrated flux, but also on the observed velocity width.  
At fixed $\s21$ the detection is more likely for
narrower HI profile widths.
\begin{eqnarray} 
\log S_{21} &=& \left\{
\begin{array}{lr}
0.5 \log W_{50} - 1.207 & : \log W_{50} < 2.5 \\
\log W_{50} - 2.457 & : \log W_{50} \geq 2.5
\end{array} 
\right.\nonumber\\
\label{eq_completeness}
\end{eqnarray}

After applying this completeness cut we are
left with a sample of 7857 galaxies. Among these, 6076 galaxies 
have spectroscopic as well as photometric measurements in SDSS DR7 and 
1633 galaxies have only photometric measurements. As for the remaining 148, 
we loosely refer to them 
as \emph{dark} galaxies. These objects are not being identified in the SDSS 
pipeline as potential galaxies. Although  follow up 
observations have been made on some of the dark galaxies
\citep{2015AJ....149...72C, 2015ApJ...801...96J, 2017ApJ...842..133L}, 
in this work we refer to all of them as dark.  This translates to around $\sim 2\%$
of galaxies which are dark or have no optical counterparts in SDSS.

The ALFALFA catalog also lists the SDSS objectIDS of the OC.
Using these we extracted the photometric properties 
like the \emph{ugriz} values (model magnitudes), 
then corrected for extinction (due to our own galaxy) \citep{1998ApJ...500..525S}
for these 7709 (non-dark) galaxies.
We also \emph{kcorrect} \citep{2007AJ....133..734B} the magnitudes 
to finally obtain rest frame magnitudes. For objects which do not 
have spectroscopic redshifts we have supplied the HI redshifts to kcorrect them.
The kcorrect code also estimates additional properties like galaxy stellar mass, 
integrated star formation history and metallicity for these objects. 

\begin{figure*}
\centering
\includegraphics[width=6.8in]{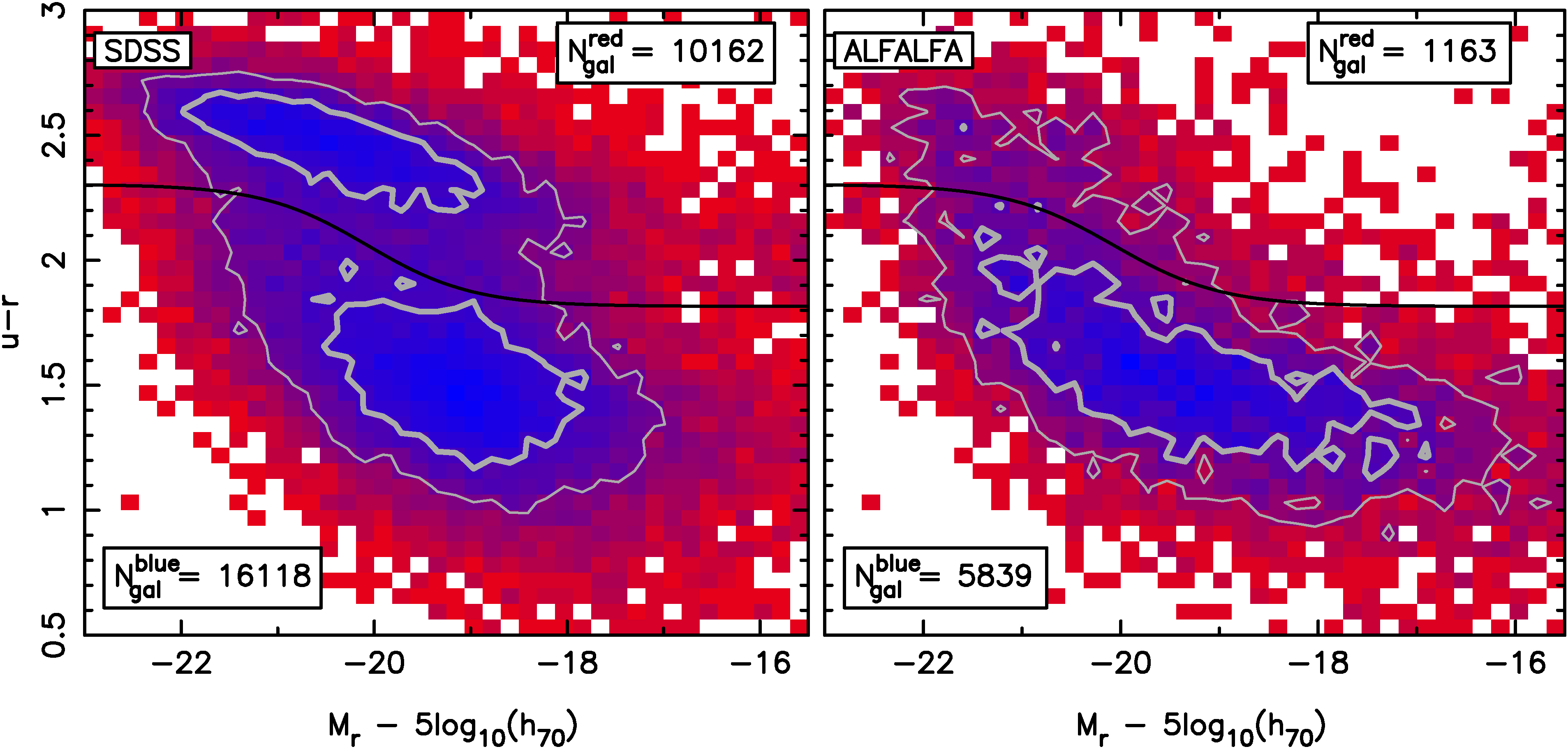}
\caption{The observed distribution of galaxies in the color-magnitude plane 
for SDSS (left panel) and ALFALFA (right panel) in a common volume considered 
in this work. The 1$\sigma$ and 0.25$\sigma$ contours  are given by the 
thin and thick lines. A double peaked, bimodal distribution of galaxies 
is visible in SDSS and
the solid curve (eq.~\ref{eq_tanh_cut}) is used to classify the galaxies into red 
(above curve) and blue (below curve) populations \citep{2004ApJ...600..681B}. The numbers
indicate the observed counts of galaxies in this color-magnitude range for 
each of the populations.}
\label{fig_sdss_alfa}
\end{figure*}

In figure~\ref{fig_sdss_alfa} we look at the observed distribution of galaxies both in SDSS
and ALFALFA in the color($u-r$)-magnitude($M_r$) plane. 
The 1$\sigma$ and 0.25$\sigma$ contours  are given by the thin and thick lines.
One can see a clear bimodality in the observed distribution in SDSS. 
The solid curve (eq.~\ref{eq_tanh_cut}) is used to classify the galaxies into red 
(above curve) and blue (below curve) populations \citep{2004ApJ...600..681B}.
The numbers indicate the observed 
counts for the red and blue populations in this color-magnitude range. 
As discussed earlier $\sim 98\%$ of ALFALFA galaxies have 
optical counterparts in SDSS. Here we show 
that ALFALFA predominantly samples the blue cloud. 
We see that about 38\%(11\%) of blue(red) galaxies in SDSS have detections in ALFALFA.

As of writing this paper the ALFALFA team has released the 100\% catalog ($\alpha.100$) 
\citep{2018ApJ...861...49H} which also include the RA and dec of the optical counterparts.
However we find that there are many galaxies which have luminous foreground stars due to 
which SDSS has masked the region covering the galaxy and photometric values 
are not provided. We therefore restrict ourselves with $\alpha.40$ catalog
and will revisit the $\alpha.100$ sample in the future.

\subsection{Subsamples and Populations of Galaxies}
\label{subsec_subsamples}
\begin{figure}
\centering
\includegraphics[width=\columnwidth]{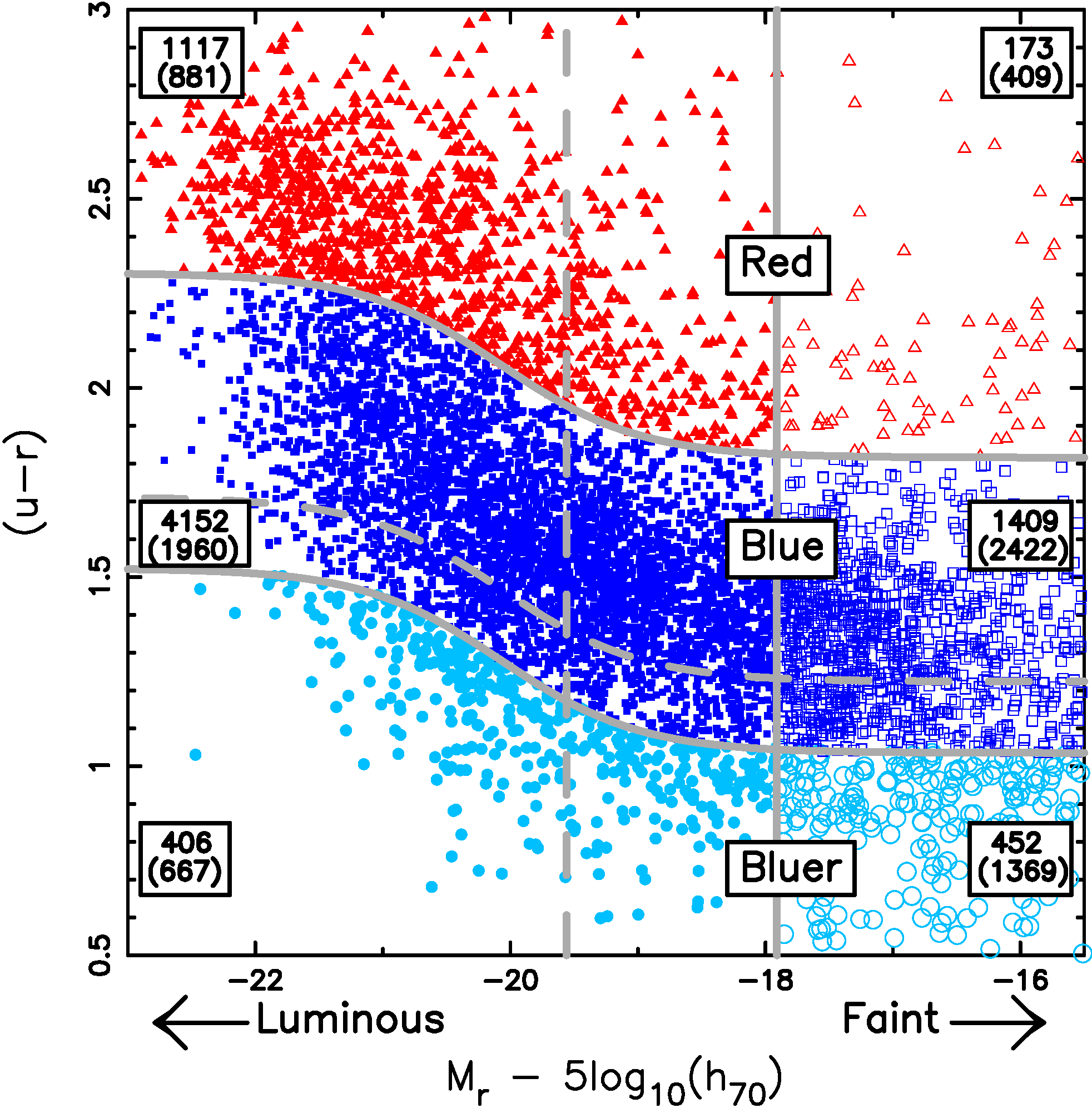}
\caption{The distribution of HI selected galaxies in this work in the color-magnitude 
plane. The upper solid curve demarcates the red (above curve) 
from the blue (below curve) population as in  \citet{2004ApJ...600..681B}. 
The vertical solid (dashed) line divides the luminous and faint 
populations.  
The line has been chosen so that the fraction of luminous
red objects over all  red objects is 0.87 which we refer as the 1.5$\sigma$ line. 
Similarly the lower solid curve is chosen so that the fraction of blue (above curve)
galaxies over all blue  
galaxies is 0.87 (or the 1.5$\sigma$ cut in color). This curve has 
been chosen to be parallel to the curve which demarcates the red and blue populations. 
We refer to the objects below(above) the curve as bluer(blue) galaxies. Similarly 
we define the 1$\sigma$ sample with the help of dashed lines. This breaks 
the sample of HI selected galaxies (which have optical counterparts) 
into 6 disjoint sets in the color-magnitude plane. 
The number for each population is quoted. The numbers in brackets are for the 
1$\sigma$ sample.}
\label{fig_colormag}
\end{figure}

From our sample of 7857 galaxies we identify subsamples which  
define different populations of galaxies.
These populations are disjoint sets and their union (including the dark galaxies)  
forms the full sample. 
The populations are based on dividing 
the color($u-r$)-magnitude($M_r$) plane of the HI selected galaxies 
into six disjoint regions. As seen in figure~\ref{fig_sdss_alfa} there are two 
distinct populations, red and blue, seen in SDSS. We will start with this definition 
to further break our full sample into different populations. 
We show this in figure~\ref{fig_colormag} (which is similar to figure~\ref{fig_sdss_alfa}), 
where we plot individual points instead of binning the data. 

In  figure~\ref{fig_colormag} the upper solid curve demarcates the red (above curve) 
from the blue (below curve) population as in  \citet{2004ApJ...600..681B}.
This optimal divider is given by 
\begin{eqnarray} 
C^\prime_{ur}(M_r) &=& 2.06 - 0.244 \tanh \left[ \frac{M_r+20.07}{1.09}\right]
\label{eq_tanh_cut}
\end{eqnarray}
The vertical solid line divides the luminous (leftward of line) and 
intrinsically faint populations (henceforth we will call these faint populations).
The line has been chosen so that the fraction of luminous
red objects over all red objects is 0.87 which we refer to as the 1.5$\sigma$ line. 
Similarly the lower solid curve is chosen so that the fraction of blue (above curve)
galaxies over all blue  galaxies is 0.87 (or the 1.5$\sigma$ cut in color). This curve has 
been chosen to be parallel to the curve which demarcates the red and blue populations. 
We refer to the objects below(above) the curve as bluer(blue) galaxies. Similarly 
we define the 1$\sigma$ sample (dashed lines). This breaks 
the sample of HI selected galaxies (which have optical counterparts) 
into 6 disjoint sets in the color-magnitude plane; we call each set a single population. 
The number for each population is quoted. The numbers in brackets are for the 
1$\sigma$ sample. The data points are: i. filled (open) triangles 
for luminous (faint) red galaxies ii. filled (open) squares for luminous (faint) 
blue galaxies iii. filled (open) circles for luminous (faint) bluer galaxies and have
been marked for the 1.5$\sigma$ sample. 

Our definition of a population of galaxies is broadly a subsample of the total sample
of 7857 HI selected galaxies. The dark population has no optical counterparts.
The luminous(faint) population is a subsample of galaxies with absolute magnitudes $M_r^i$,
such that $M_r^i < M_r$ ($M_r^i \geq M_r$) for some reference value $M_r$ (the vertical line
in figure~\ref{fig_colormag}). Similarly we have defined three populations 
based on color (red, blue and bluer), based on equation~\ref{eq_tanh_cut}. The six 
populations are formed by further splitting these three into luminous and faint populations.
As discussed earlier, each population is dependent 
on the boundaries that define them which we have called a 1$\sigma$ or 1.5$\sigma$ line.
We caution the reader that our definition of sample may be confusing  since 
both these samples represent the same set of 7709  galaxies which have optical 
counterparts. However the number of galaxies for any given population 
is different for the $1\sigma$ and $1.5\sigma$ samples.    
For the $1.5\sigma$ sample the fraction of detections
of ALFALFA with respect to SDSS for the luminous red, blue and bluer populations
are 11\%, 32\%, 39\% and it is 14\%, 58\%, 62\% for the faint populations.

We have chosen to define two samples to demonstrate that the qualitative results do not 
depend very strongly on sample definition. However as we will show in the next section,
the definition of the faint bluer population (i.e. the 1.5$\sigma$ or 1$\sigma$ sample) 
determines whether it contributes significantly or not to some part of the HIMF.
Looking at how the observed counts change,
both in relative and absolute terms,  when going from the 1.5$\sigma$ to the 1$\sigma$
sample we expect that the HIMF of the 
luminous blue population to be the affected the most. 
In section~\ref{sec_discussion} (figure~\ref{fig_mstar-mh1}) we will discuss 
the break in the $M_{\text{HI}}-M_{\text{star}}$,  at $M_{\text{star}} \simeq 9$, 
where $M_{\text{star}}$ is the logarithm of the galaxy stellar mass in units of 
$\msun/h^{2}_{70}$.  
One can then use the mean $M_r-M_{\text{star}}$ relation to convert  $M_{\text{star}} = 9$ to 
$M_r = -19$. This value is sandwiched between the vertical lines (solid and dashed) 
in figure~\ref{fig_colormag}.

\section{Results}
\label{sec_results}
The mass function $\phi(M)$ is defined as 
the number density of objects in the mass range $[M,M+dM]$.
The HI mass function $\phi (M_{\text{HI}})$ 
can be expressed as 
\begin{eqnarray}
\phi (M_{\text{HI}}) &=& \frac{dN}{V dM_{\text{HI}}}
\end{eqnarray}
where, $dN$ is the total number of galaxies
having HI mass between $M_{\text{HI}}$ and 
$M_{\text{HI}}$+$dM_{\text{HI}}$ and
$V$ is the survey volume of interest.
The HI mass function can be parameterized as a Schechter Function
\begin{eqnarray}
\phi (M_{\text{HI}}) &=& \phi_* \left( \frac{M_{\text{HI}}}{M_*}\right)^{\alpha} \exp \left(- \frac{M_{\text{HI}}}{M_*}\right)
\label{eq_schecterfn}
\end{eqnarray}
Here, $\alpha$ is the faint end slope, 
$\phi_*$ is the amplitude and $M^*$ is the characteristic HI mass.

A simple and intuitive way of calculating the HIMF is by the 
 1/$\vmax$ method \citep{1968ApJ...151..393S}.
The underlying assumption in this method is that the sample 
of galaxies detected by a survey is a representative sample 
of  galaxies in the Universe (in the same redshift range),
or in other words it assumes homogeneity. For each detected galaxy 
'\emph{i}' a maximum detectable distance $\dmax^i$ is calculated based on 
inverting eq.~\ref{eq_massfluxrelation} by using the  mass $\mhi^i$ 
of the galaxy and the limiting flux $S_{\text{lim}}^i$ which is the property of the survey. 
In the case of ALFALFA the completeness relation (eq.~\ref{eq_completeness}) 
determines this limiting flux for the velocity width $\w50^i$ of the galaxy.
$\dmax^i$ is then converted to a volume $\vmax^i$ which is the volume 
in which the galaxy could be in, and still be detected by the survey. 
Finally the galaxies are binned in mass with relative weights $1/\vmax^i$ 
to obtain the mass function. 
For the more luminous or more massive objects  $\vmax^i$ is larger 
than the survey volume. In such a case the relative weight is set to unity. 
This method has the advantage that it is non-parametric and does not require any prior 
knowledge to estimate the HIMF. However since galaxies cluster,  
the estimate of the HIMF will be sensitive to large scale structure in the local volume.
Additionally some volumes of the survey may be inaccessible due to RFI 
and needs further correction. An estimate of these effects can be used 
to recalibrate the weights $\vmax^i$ \citep{2010ApJ...723.1359M}.

Maximum likelihood \citep{1979ApJ...232..352S} and Step-Wise Maximum Likelihood (SWML) 
\citep{1988MNRAS.232..431E} methods on the other hand are designed to 
be insensitive to large-scale structure. In the former the assumption is that the galaxy 
sample is drawn from an underlying distribution function, e.g. the HIMF for this work 
$\phi(\mhi)$, and the likelihood method determines the parameters of this function. 
Although in most cases a Schechter function is the chosen function one has no way of testing
whether it is the optimal function to describe data \citep{1988MNRAS.232..431E}. \citet{2012MNRAS.421..621B} and \citet{2009ApJ...707.1595D} find that a single Schechter function 
does not describe the galaxy stellar mass function. In the latter, i.e. SWML, 
the underlying distribution $\phi(\mhi)$ has no functional form but is discretized in 
steps or bins of mass and a uniform distribution is assumed in each bin. Hence the value 
$\phi_j$ which is the value of $\phi$ in the $j^{\text{th}}$ mass bin becomes the parameter. 
The joint likelihood of detecting all galaxies 
in the sample is maximized, with respect to the parameters $\phi_j$, thus determining their 
values. This method works when the sample is flux-limited. 

When the selection function depends on other properties of the galaxies
one needs to consider an underlying bivariate or multivariate 
distribution for $\phi$. One has to then generalize the SWML method
to higher dimensions. \citet{2000MNRAS.312..557L} estimated 
the bivariate luminosity function $\phi(M_\text{K},M_\text{B})$ and 
then marginalized over $M_\text{B}$ to obtain the K-Band luminosity 
function, $\phi(M_\text{K})$, 
starting with a $b_{J}$-selected sample in the Stromlo-APM Redshift Survey.

For a blind HI survey like ALFALFA the limiting flux 
(figure~\ref{fig_sw} and eq.~\ref{eq_completeness}) 
depends on the velocity width $\w50$. A two-dimensional SWML (2DSWML) method similar 
to \citet{2000MNRAS.312..557L} was applied by \citet{2003AJ....125.2842Z}
to estimate the HIMF for HIPASS Galaxies. The bivariate distribution
in this case is $\phi(\mhi,\w50)$ which can be marginalized over $\w50$ to obtain 
the HIMF \citep{2003AJ....125.2842Z, 2010ApJ...723.1359M, 
2011AJ....142..170H, 2018MNRAS.477....2J} 
or marginalized over $\mhi$ to obtain the HI velocity width function 
\citep{2010MNRAS.403.1969Z, 2014MNRAS.444.3559M}.
The details of our implementation are given in appendix~\ref{appendix}.

\begin{figure*}
  \begin{tabular}{cc}
    \includegraphics[width=3.4in]{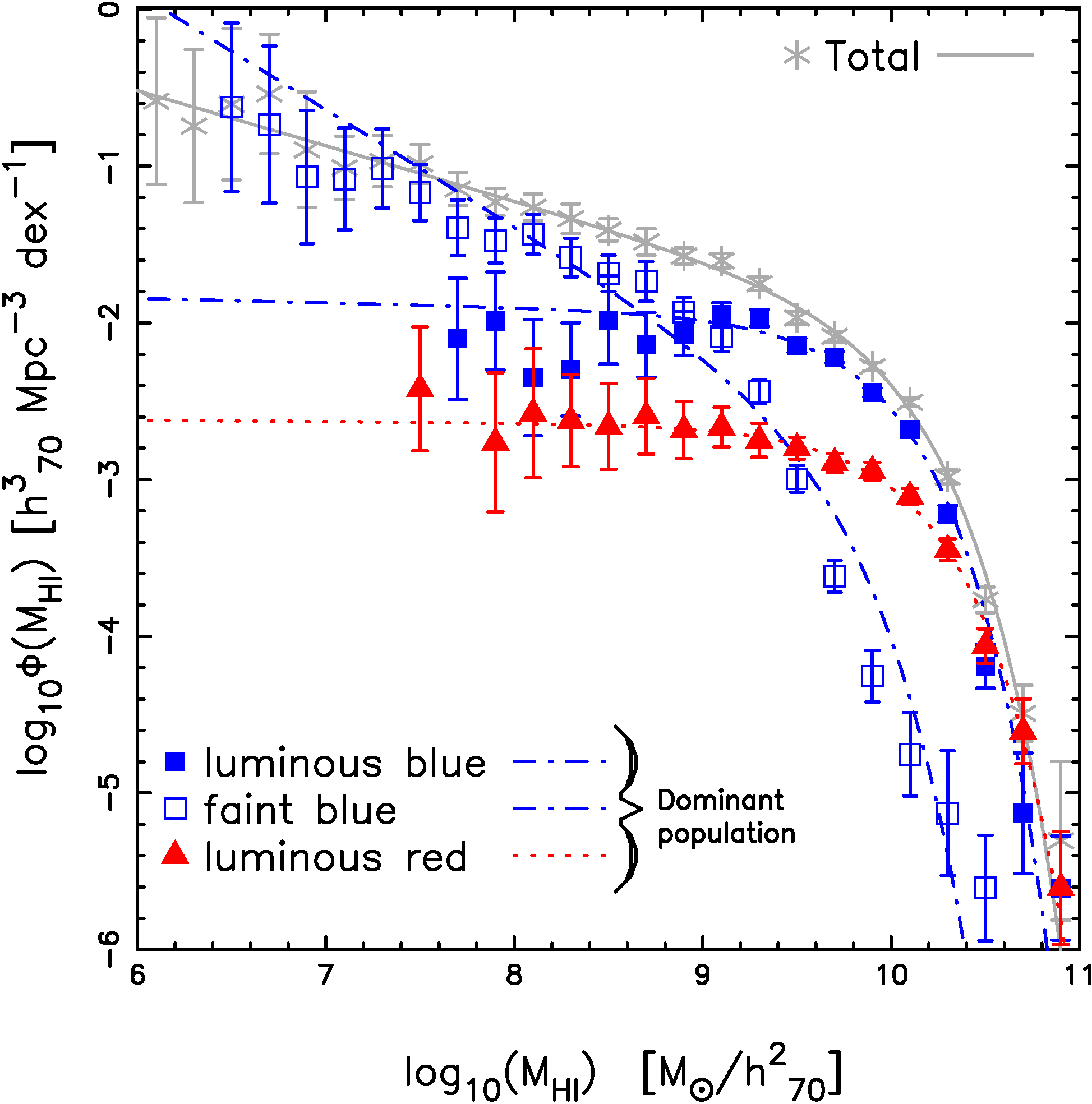}
    \includegraphics[width=3.4in]{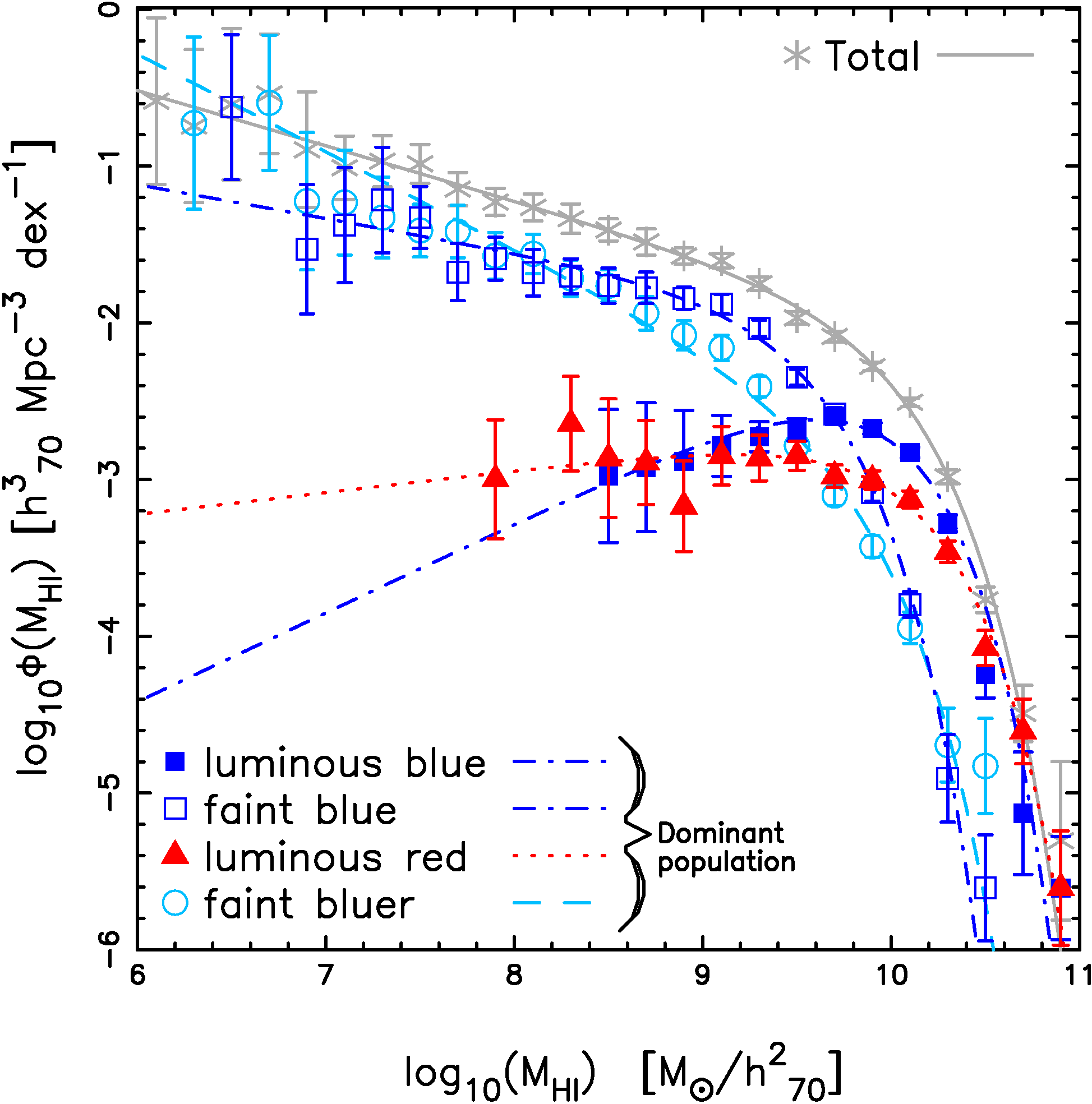}\\
    \includegraphics[width=3.4in]{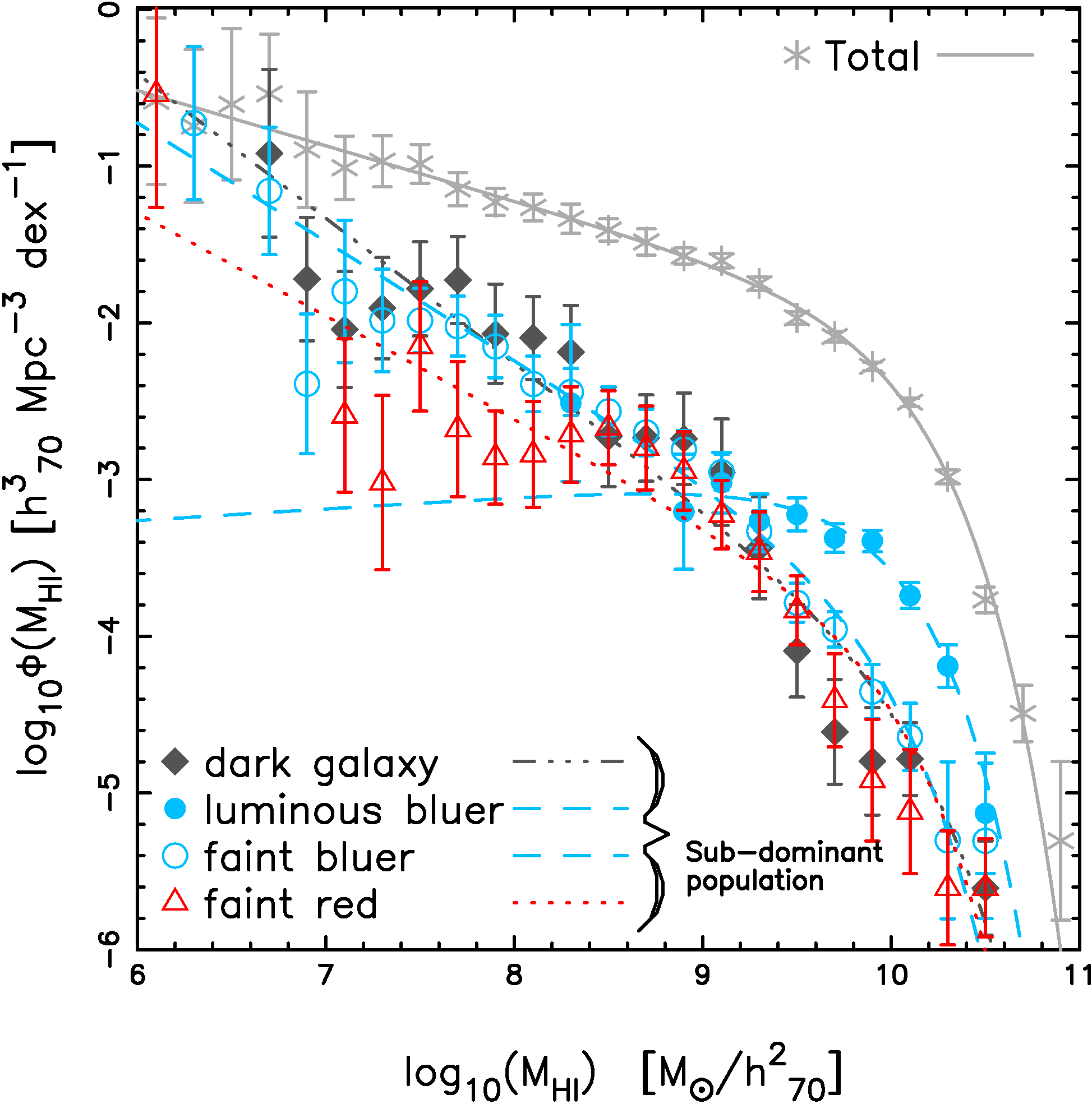}
    \includegraphics[width=3.4in]{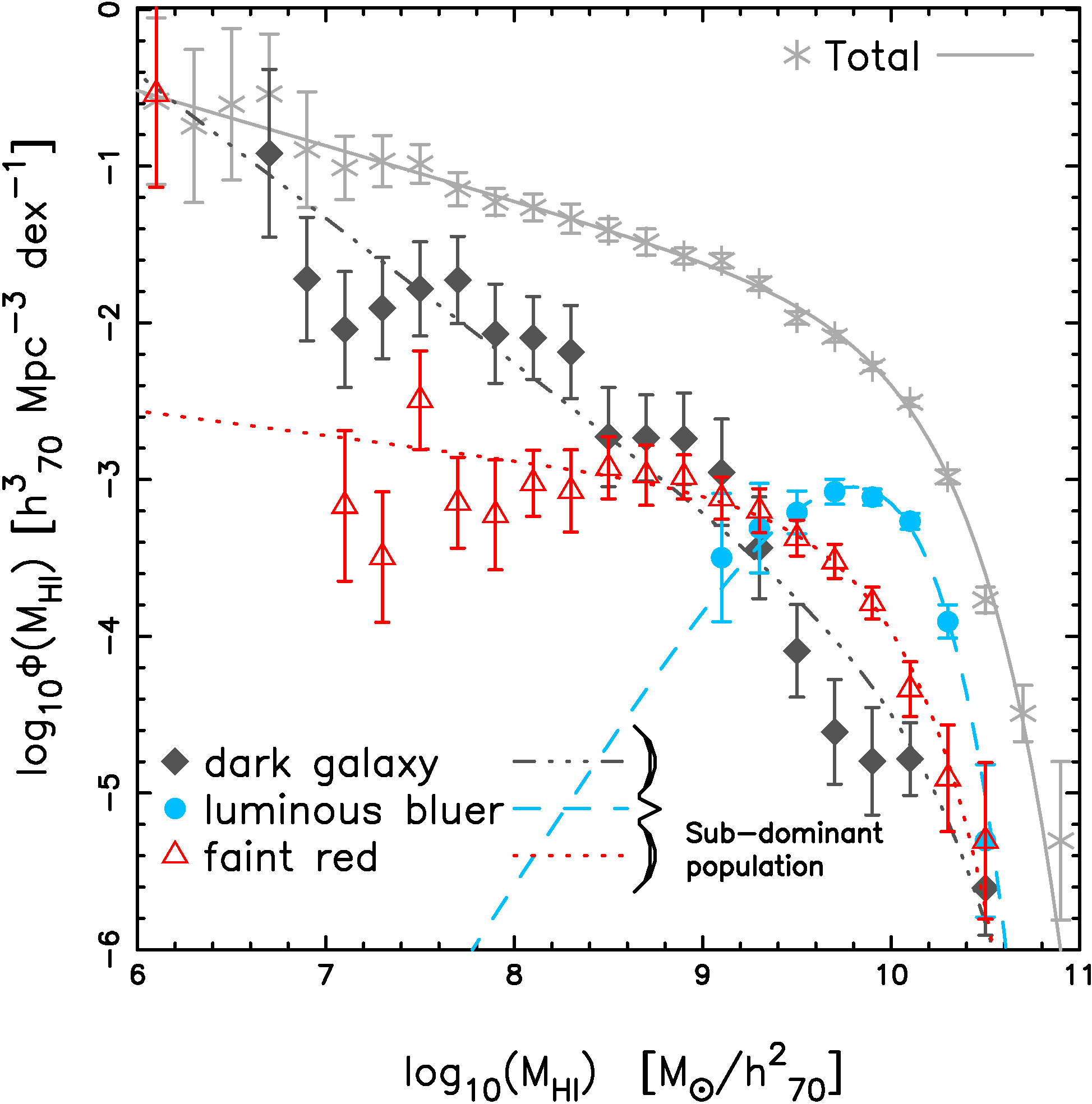}\\
  \end{tabular}
  \caption{The HIMF for the $1.5\sigma$ (left column) and the $1\sigma$ (right column) 
    samples. Data points and error bars were estimated using the 2DSWML method 
    (see appendix~\ref{appendix}). The curves are Schechter function fits.
    For comparison we have added the total HIMF (crosses) and its fit (solid line)
    in all the panels. For each sample the top row is for the dominant population
    and the bottom for the sub-dominant one. In addition to the six 
    populations we have also 
    added the HIMF for the dark population (filled diamonds) and its Schechter function fit
    (dot-dot-dot-dashed line). The symbols 
    for the six populations are the same as in figure~\ref{fig_colormag}. The Schechter fits
    are given for the red (dotted), blue (dot-dashed) and bluer (dashed) populations. 
    The details of the fits are given in table~\ref{tab_schecterfit}.}
  \label{fig_himf}
\end{figure*}

In figure~\ref{fig_himf} we show our estimate of the HIMF for all the  populations
including the total (crosses) and the dark (filled diamonds) populations. 
The columns are for the 
$1.5\sigma$ (left) and the $1\sigma$ (right) populations. We have 
broken our results into 
a dominant (top row) and a subdominant (bottom row) population to better 
illustrate our results. We call a dominant population one which dominates the HIMF 
over the rest of the populations in some mass range and also contributes greater than 
10\% to $\Omega_{\text{HI}}^{\text{tot}}$ (see table~\ref{tab_omegah1}). 
The symbols for the six populations are the same as in figure~\ref{fig_colormag}. 
The curves are Schechter function fits for the total (solid), red (dotted), 
blue (dot-dashed), bluer (dashed) and dark (dot-dot-dot-dashed) populations.
Our Schechter function fits are summarized in table~\ref{tab_schecterfit}.
In the rest of the paper we will quote the values of the characteristic 
mass $M_*$ and the amplitude of the Schechter function $\phi_*$ 
in the units $\log (M_*/M_{\odot}) + 2\log h_{70}$ and  $(10^{-3} h_{70}^{3} Mpc^{-3} dex^{-1})$
respectively. We will also quote the values of $M_{\text{HI}}$ in the same units as $M_*$.
In practice we bin the mass and the velocity width in logarithmic bins
therefore the faint slope in figure~\ref{fig_himf} ($\alpha^\prime$) differs 
from $\alpha$ in equation~\ref{eq_schecterfn} by 1, or $\alpha^\prime = \alpha + 1$.  
 
\begin{table*}
\begin{center}
\begin{tabular}{l|c|c|c|c}
\hline
	region & $\log (M_*/M_{\odot}) + 2\log h_{70}$ & $\phi_*$ & $\alpha$ & $\chi^2_{reduced}$ \\
	& & $(10^{-3} h_{70}^{3} Mpc^{-3} dex^{-1})$ & &  \\
	\hline
	total & 9.96 $\pm$ 0.02 & 5.34 $\pm$ 0.40 & -1.35 $\pm$ 0.02 & 0.79 \\
	\hline
	luminous blue & 9.86 $\pm$ 0.02 (9.85 $\pm$ 0.03) & 4.85 $\pm$ 0.42 (2.54 $\pm$ 0.17) & -1.03 $\pm$ 0.06 (-0.43 $\pm$ 0.11) & 2.02 (2.17) \\
	\hline
	faint blue & 9.57 $\pm$ 0.04 (9.50 $\pm$ 0.02) & 1.22 $\pm$ 0.22 (5.84 $\pm$ 0.58) &  -1.74 $\pm$ 0.05 (-1.22 $\pm$ 0.05) & 4.10 (0.96) \\
	\hline
	luminous red & 10.04 $\pm$ 0.04 (10.02 $\pm$ 0.04) & 0.96 $\pm$ 0.12 (0.95 $\pm$ 0.12) & -1.01 $\pm$ 0.08 (-0.86 $\pm$ 0.10) & 0.32 (0.52) \\
	\hline
	luminous bluer & 9.84 $\pm$ 0.08 (9.52 $\pm$ 0.05) & 0.46 $\pm$ 0.13 (0.79 $\pm$ 0.16) & -0.92 $\pm$ 0.23 (0.86 $\pm$ 0.38) & 0.64 (0.38) \\
	\hline
	faint bluer & 9.83 $\pm$ 0.07 (9.72 $\pm$ 0.04) & 0.10 $\pm$ 0.03 (1.09 $\pm$ 0.18) &  -1.76 $\pm$ 0.06 (-1.62 $\pm$ 0.05) & 1.11 (1.40) \\
	\hline
	faint red & 9.89 $\pm$ 0.07 (9.74 $\pm$ 0.06) & [6.09 $\pm$ 2.07] $\times 10^{-2}$ (0.31 $\pm$ 0.07) & -1.66 $\pm$ 0.11 (-1.16 $\pm$ 0.08) &  1.27 (1.36) \\
	\hline
	dark & 10.03 $\pm$ 0.09 & [3.25 $\pm$ 1.50]$\times 10^{-2}$ & -1.92 $\pm$ 0.09 & 1.17 \\
	\hline
\end{tabular}
\end{center}
\caption{Parameters of the Schechter function fit to the HIMF for all the populations. 
  The estimated parameters and their uncertainties are for the $1.5\sigma$ sample 
  and the numbers in brackets are for the $1\sigma$ sample.The goodness of fit, $\chi^2_{reduced}$ is given in the last column. }
\label{tab_schecterfit}
\end{table*}

Since we are working with Code 1 objects in the $\alpha$.40 sample it is appropriate 
to compare the total HIMF to that of \cite{2010ApJ...723.1359M}. For the  HIMF 
we find that our results of ($M_* \pm \sigma_{_{M_*}},\phi_* \pm \sigma_{\phi_*}, 
\alpha  \pm \sigma_{\alpha}$) = 
($9.96\pm 0.02, 5.34 \pm 0.40, -1.35 \pm 0.02$) and $\chi^2_{reduced} = 0.79$ 
(table~\ref{tab_schecterfit}). This is consistent at the $1\sigma$ 
level with \cite{2010ApJ...723.1359M} who found 
($M_* \pm \sigma_{_{M_*}},\phi_* \pm \sigma_{\phi_*}, 
\alpha  \pm \sigma_{\alpha}$) = 
($9.96\pm 0.02, 4.8 \pm 0.30, -1.33 \pm 0.02$). Note however that $\phi_*$ 
is barely within $1\sigma$ of each other. We attribute this difference to the choice 
of sample in this work which has $\sim 25\%$ fewer galaxies 
than  \cite{2010ApJ...723.1359M}. We point out that when we consider the full sample
our results match well (see e.g. figure~\ref{fig_alfa100MF}).

\begin{figure}
\centering
\includegraphics[width=\columnwidth]{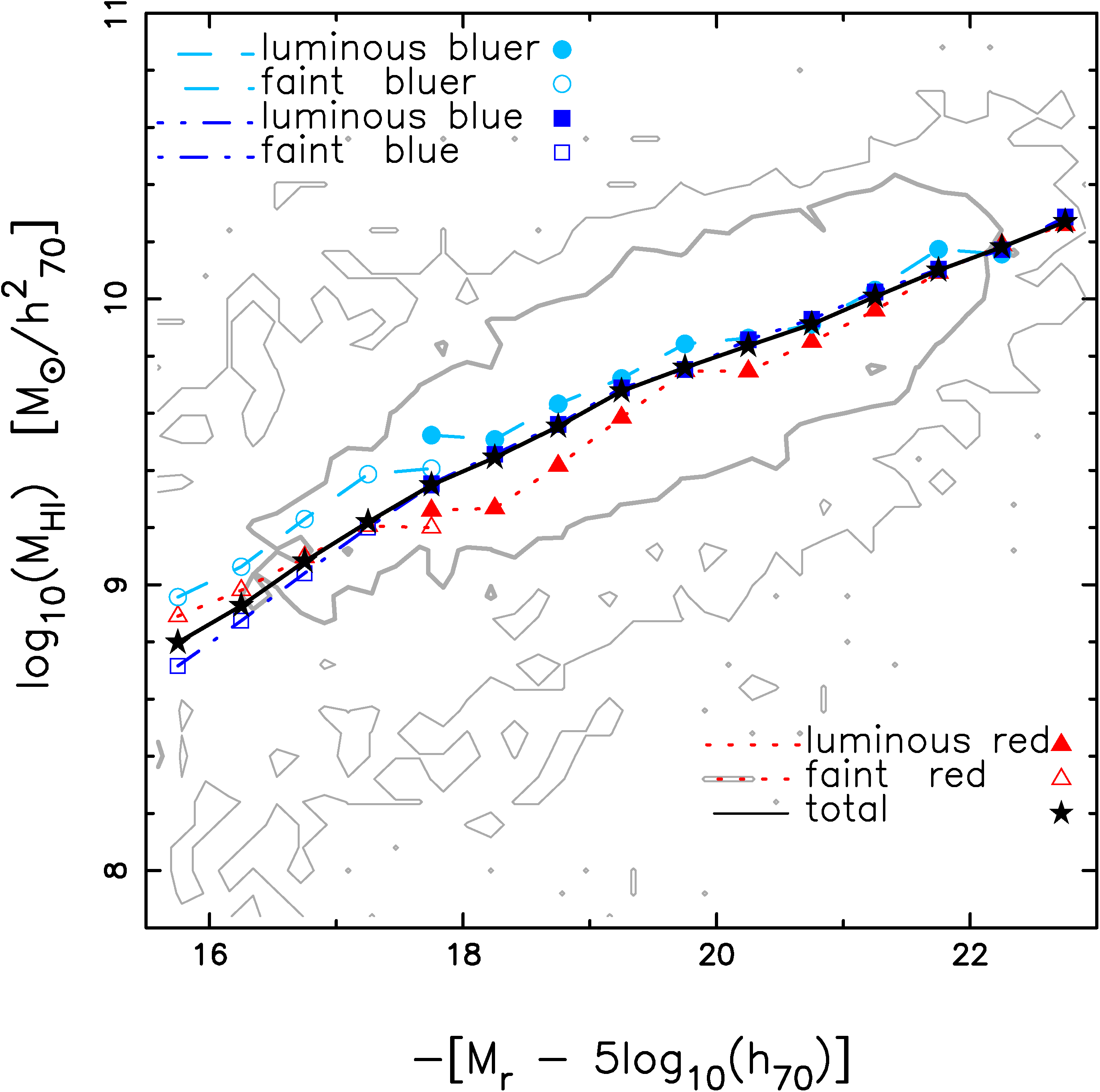}
\caption{The $M_{\text{r}}-M_{\text{HI}}$ relation for the $1.5\sigma$ sample. 
  The filled star represents the total sample excluding the dark galaxies. The 
  other data points and line styles are the same as in figure~\ref{fig_himf}.}
\label{fig_mr-mh1}
\end{figure}

The goodness of fits, $\chi^2_{reduced}$, is given in the last column of 
table~\ref{tab_schecterfit}. 
For the dark, faint red, faint bluer populations the  $\chi^2_{reduced}$ is of order unity.
In the $1\sigma$ sample of the faint red population the biggest contribution of 
$\chi^2_{reduced}$ comes from the lowest mass bin. If we remove that point as an outlier 
then $\chi^2_{reduced} = 0.57$ and $\alpha$ flattens to $-1.06$ from $-1.16$, 
$\phi_*$ increases by about $22\%$ from $0.31$. The change in $M_*$ is negligible. 
The luminous red and luminous bluer populations have a low $\chi^2_{reduced}$, irrespective of 
sample definition, and looking at the data points relative to their fitted curve 
we find that there is little variation between them. This means that the error bars 
are larger than the variation between data and model. The luminous blue and faint blue 
populations on the other hand have  larger $\chi^2_{reduced}$. For the $1.5\sigma$ sample 
of the faint blue population   $\chi^2_{reduced} = 4.10$ and improves  to 
 $\chi^2_{reduced} = 0.96$ for  the $1\sigma$ sample. The $1\sigma$ sample
for the faint blue population has a larger number of luminous galaxies 
as compared to the $1.5\sigma$ sample (figure~\ref{fig_colormag}) 
and because $M_{\text{HI}}$ is correlated to $M_r$,
(see figure~\ref{fig_mr-mh1}) the high mass end of the HIMF
is better represented in the $1\sigma$ sample, leading to a smoother change in data
and a better fit. This can also be 
seen in the top row of figure~\ref{fig_himf}. The luminous blue population 
has  $\chi^2_{reduced} \sim 2$ irrespective of sample definition.

We start by looking at the luminous population.
Due to the monotonic relation between $M_{\text{HI}}$ and $M_r$ 
(figure~\ref{fig_mr-mh1}), across populations,  
we expect the luminous population to dominate the massive end of the HIMF
and be subdominant at the low mass end. This is seen in figure~\ref{fig_himf}.
The characteristic mass, $M_*$ increases systematically  from the luminous bluer to 
the luminous red population. There is little change in  $M_*$ for both the luminous red
and luminous blue populations with respect to sample definition. 
We also see little change in the HIMF
with respect to sample definition for $M_{\text{HI}} \geq 10.3$ for both these populations.  
The luminous bluer population on the other hand has $M_* = 9.84$ for the $1.5\sigma$ sample
and reduces to $M_* = 9.52$ for the $1\sigma$ sample. 
At the low mass end the luminous populations
have shallower slopes $\alpha + 1 \geq 0$. 
This is expected as mentioned earlier since at the low 
mass end we expect the faint population to dominate. At this end, 
the sample definition affects 
the luminous bluer population the most, where $\alpha = -0.92$ ($1.5\sigma$ sample)
and increases to   $\alpha = +0.86$ ($1\sigma$ sample), the change being the smallest 
for the luminous red population. On the other hand the amplitude is most affected 
for the luminous blue population, it changes from $\phi_* = 4.85$  ($1.5\sigma$ sample)
to $\phi_* = 2.54$  ($1\sigma$ sample). The change is negligible for the luminous red 
and about $\sim 72\%$ for the luminous bluer population. We point out that  changes
in the HIMF with respect to sample definition 
can be best understood in terms of how the observed number of galaxy populations change
when the boundaries in the color-magnitude plane are redrawn to define
a new sample (figures~\ref{fig_colormag}) and 
the average scaling relation $M_{\text{r}}-M_{\text{HI}}$ (figure~\ref{fig_mr-mh1}).  
For the $1.5\sigma$ sample 
the luminous blue is the dominant population at the knee of the HIMF, however 
it contributes nearly equally as the faint blue population when we consider the 
$1\sigma$ sample. This is because  the faint blue population has a net 
increase in observed galaxies from $n_{\text{gal}} = 1409$ ($1.5\sigma$ sample) to 
$n_{\text{gal}} = 2422$ ($1\sigma$ sample), the net change coming from the intersection 
of luminous blue ($1.5\sigma$) and faint blue ($1\sigma$) (see figure~\ref{fig_colormag});
this is the primary reason for the increase of $\phi_*$ by  $\sim 4.8\times$.
An interesting result is that the luminous red population is the dominant 
population at $M_{\text{HI}} \geq 10.4$. This result is insensitive to sample definition.
We find that for $M_{\text{HI}} \geq 10.4$ the luminous red population 
represents $\sim 60\%$ of total detections 
with $\sim 40\%$ coming from the luminous blue population 
which also translates to similar
fractions in total HI mass at this end.

We now move to the faint population. All of them have steeper slopes as compared 
to their luminous counterparts and do not dominate the HIMF at the high mass end. 
The faint red population is the most subdominant population. The observed 
counts of galaxies of the faint
bluer population see the largest relative change with sample definition increasing
from $n_{\text{gal}} = 452$ ($1.5\sigma$ sample) to 
$n_{\text{gal}} = 1369$ ($1\sigma$ sample), a factor $\sim 3\times$. This results 
in a small change in slope from $\alpha = -1.76$ to $\alpha = -1.62$ but a large,  
 $\sim 10\times$, increase in amplitude from $\phi_* = 0.10$ to   
$\phi_* = 1.09$. In the $1.5\sigma$ sample the faint bluer population is a subdominant 
population, however it becomes the dominant population for masses $M_{\text{HI}} \leq 8.1$.
The faint blue population is on the other hand a dominant population 
below the knee of the mass function. It dominates the HIMF for  
 $M_{\text{HI}} \leq 8.7$ in the $1.5\sigma$ sample and for  
$ 8.1 \leq M_{\text{HI}} \leq 9.7 $ in the $1\sigma$ sample.

The dark population is characterized by a very steep slope $\alpha = -1.92$, 
large characteristic mass $M_* = 10.03$ and a very small amplitude 
$\phi_* = 3.25 \times 10^{-2}$ and is a subdominant population. Extrapolating to 
masses below $M_{\text{HI}} \leq 6$ our results suggest the dark population 
will be the dominant population. However it is unclear how far down we can extrapolate 
since it is unlikely that there will be too many low mass, gas rich galaxies devoid 
of stars which will be able to self shield themselves from the photoionizing background.

\subsection{The contribution of different galaxy populations to $\Omega_{\text{HI}}$}

We can analytically integrate the HIMF, 
fitted to a Schechter function to obtain the cosmic HI density
parameter, 
\beq
\Omega_{\text{HI}} = \frac{\rho_{\text{HI}}}{\rho_c} = 
\frac{M_* \phi_*}{\rho_c}\Gamma(\alpha+2)
\label{eq_omegah1}
\eeq
Alternately we can sum the binned measurements of the HIMF.
Similar to \citet{2011AJ....142..170H} 
we find that both methods give similar results, with the exception 
of the dark sample which has a very steep slope. As we argued in 
the previous section it is not physical to extrapolate the HIMF to very small 
masses. Hence we choose to quote our results by the \emph{summed} method.
We note that the results do not change if we integrate the Schechter function 
from $M_{\text{HI}} = 6.1$ to $\infty$.
We summarize our results in table~\ref{tab_omegah1}. Column 2 is the estimate of 
$\Omega_{\text{HI}}$ from each population and column 3 is the fractional contribution 
to $\Omega_{\text{HI}}^{\text{tot}}$, the values in brackets are for the $1\sigma$ sample.

For the total sample we obtain 
$\Omega_{\text{HI}} = (4.894 \pm 0.469) \times 10^{-4} h_{70}^{-1}$ which is consistent 
at $1\sigma$ with $\Omega_{\text{HI}} = (4.4 \pm 0.1) \times 10^{-4} h_{70}^{-1}$ (summed) 
of \cite{2010ApJ...723.1359M}. With the addition of Code 2 objects in the 
$\alpha$.40 sample  
we see that it is only consistent at the 2$\sigma$ level with \cite{2011AJ....142..170H}
who obtain  $\Omega_{\text{HI}} = (4.2 \pm 0.1) \times 10^{-4} h_{70}^{-1}$ (summed).
Since our $M_*$ is comparable with the $\alpha$.40 results, and $\alpha$ is only 
a bit steeper, the main reason for this discrepancy can be traced to $\phi_*$ 
(equation~\ref{eq_omegah1}). Our value of $\phi_*$ is $\sim 10\%$ higher than 
\cite{2010ApJ...723.1359M} which translates to a $10\%$  higher estimate of 
$\Omega_{\text{HI}}$ at fixed $M_*$ and $\alpha$. However the relative ratios
should not be sensitive to this change.

From table~\ref{tab_omegah1} we see that the red population (luminous and faint) 
have a non-negligible contribution of $\sim 17\%$ to $\Omega_{\text{HI}}^{\text{tot}}$.
When combined with the dark population ($3\%$ of  $\Omega_{\text{HI}}^{\text{tot}}$),
this adds up to a non-negligible fraction of 20\% (rounded).  
The full blue population (faint and luminous blue and bluer) then contributes 
$\sim 80\%$ (rounded) 
of  $\Omega_{\text{HI}}^{\text{tot}}$. We will discuss the implications of these numbers 
in the next section. The dominant sample about the knee of the HIMF 
are  the luminous blue 
and the faint blue populations (as $M_*$ galaxies) and together they 
contribute most to   $\Omega_{\text{HI}}^{\text{tot}}$, $\sim 73\%$ ($1.5\sigma$ sample) 
and $\sim 55\%$ ($1\sigma$ sample)

In section~\ref{sec_results} we defined a dominant population as one which dominates
the total HIMF in some mass range and contributes more than 10\% to 
$\Omega_{\text{HI}}^{\text{tot}}$. For the $1.5\sigma$ sample the dominant 
populations are the luminous red, luminous blue and faint blue populations. Whereas 
for the $1\sigma$ sample the dominant populations are the luminous red, luminous blue,
faint blue and faint bluer populations. The dominant populations together contribute 
about 90\% to $\Omega_{\text{HI}}^{\text{tot}}$, irrespective of sample definition. 
By integrating the HIMF for the dominant populations we find that they represent 
about 85\% (90\%) of galaxies above $M_{\text{HI}} \geq 8$ ($M_{\text{HI}} \geq 9$) for the 
$1.5\sigma$ sample. The numbers are similar for the $1\sigma$ sample. We do not present
numbers for a lower mass threshold (which would dominate the total number density) since 
all populations do not have detections at lower masses
and the errorbars for the HIMF are considerably large at lower masses.

\section{Discussion}
\label{sec_discussion}

\begin{figure}
\centering
\includegraphics[width=\columnwidth]{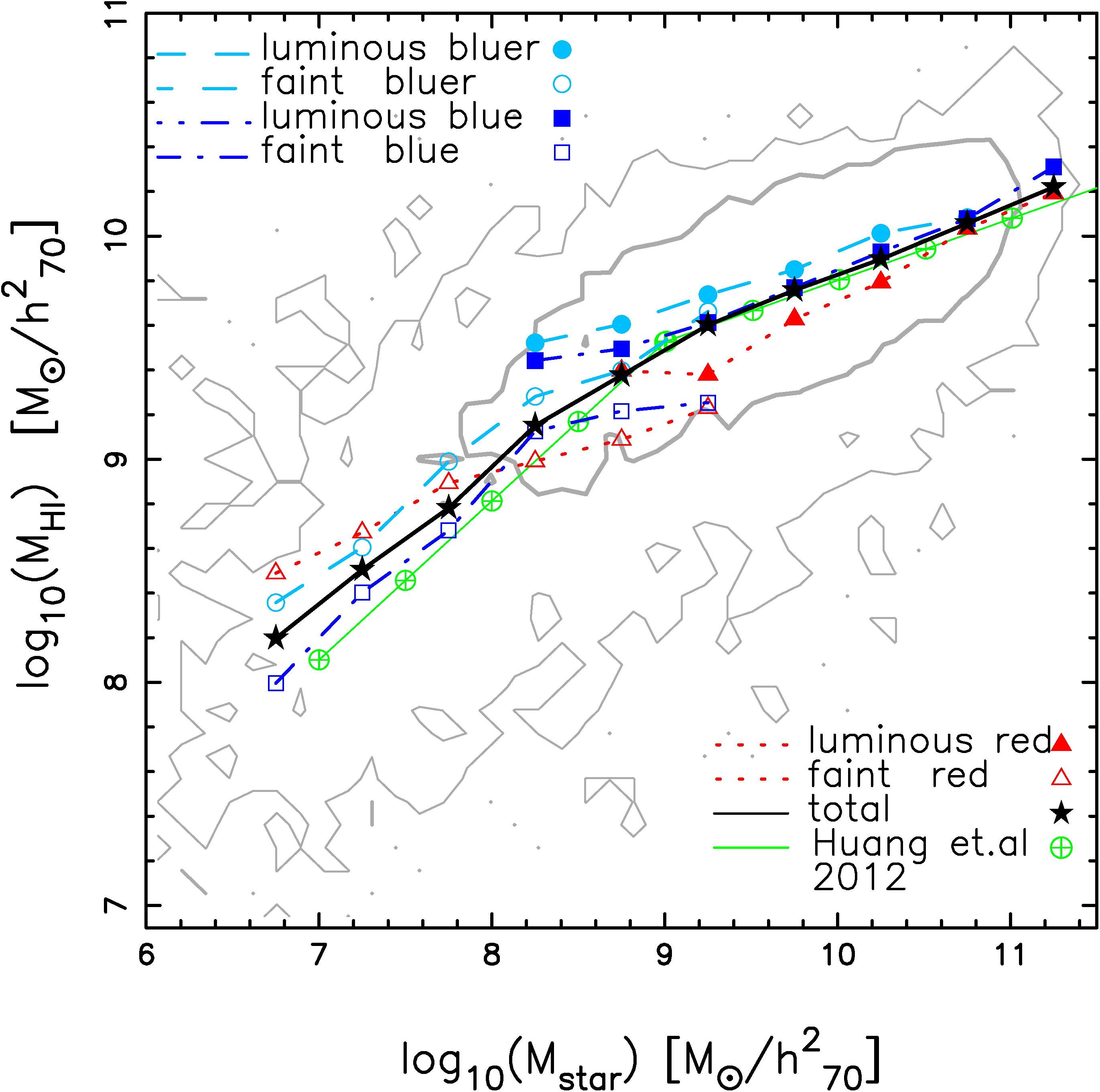}
\caption{The $M_{star}-M_{\text{HI}}$ relation for the $1.5\sigma$ sample. 
  The filled star (thick solid line) 
  represents the total sample excluding the dark galaxies. 
  The results are compared with  \citet{2012ApJ...756..113H} 
  (crossed-circle and thin solid line). 
  other data points and line styles are the same as in figure~\ref{fig_himf}.}
\label{fig_mstar-mh1}
\end{figure}

Since we are looking at the contribution of different galaxy  populations 
to the total HIMF we would like to see whether 
these populations have different scaling relations, e.g. in the 
$M_{\text{HI}}-M_{\text{star}}$ plane. Such relations have been explored 
for galaxies in the ALFALFA sample
\citep{2010MNRAS.403..683C,2012ApJ...756..113H,2015MNRAS.447.1610M} and the 
HI Parkes All-Sky Survey Catalog (HICAT) \citep{2018ApJ...864...40P}. 
In this work the stellar masses are estimated by kcorrect 
which uses the population synthesis code of \cite{2003MNRAS.344.1000B}.
Our estimates on $M_{\text{star}}$ differ from the estimates of these authors.
The $M_{\text{HI}}-M_{\text{star}}$ scaling relations are shown in figure~\ref{fig_mstar-mh1}.

In order to avoid crowding figure~\ref{fig_mstar-mh1} we choose 
to compare our results for the total sample with 
\cite{2012ApJ...756..113H} (crossed-circle, thin solid line) only. 
The $1\sigma$ scatter on the data points (total) is $\sim 0.5$ dex. 
For the total sample we find that our results compare well (within the scatter) 
with \cite{2012ApJ...756..113H} in figure~\ref{fig_mstar-mh1}.
The scaling relations that we find are also consistent with 
\cite{2015MNRAS.447.1610M,2018ApJ...864...40P} 
($M_{\text{HI}}$ selected sample) but differ from 
\cite{2010MNRAS.403..683C} ($M_{\text{star}}$ selected sample).
However our stellar masses are underestimated at lower masses. This difference can be 
attributed to the choice of sample but more so due to attenuation by dust, affecting the 
redder sample, which these authors have considered. In this work we have not attempted 
to correct for 
reddening due to dust while \cite{2012ApJ...756..113H} have used the additional two UV 
bands in GALEX to correct for it. 
Not correcting for it should therefore change the average scaling relations. 
This is also evident when looking 
at the scaling relations for the three faint populations. 
The faint blue and 
faint bluer populations have similar slopes but 
these are steeper compared to their corresponding
luminous populations. 
The faint red population, on the other hand, 
has a shallower slope with respect to the luminous 
red population as well as the faint blue and bluer populations. 
We also find that the HI fraction, $f_{\text{HI}} = M_{\text{HI}}/M_{\text{star}}$, 
increases with decreasing (\emph{u-r}) color.

We see a clear transition in the scaling relations when going from the low mass 
to the high mass end. $f_{\text{HI}}$ gets suppressed for the total sample 
at about $M_{\text{star}} \sim 9$ consistent with 
\cite{2012ApJ...756..113H,2015MNRAS.447.1610M}. The transition scale also 
depends on the galaxy population. For the blue, bluer and red populations
it occurs at $M_{\text{star}} = 10.1, 9.4, 8$ respectively. The transition scale 
 of $M_{\text{star}} \sim 9$ corresponds to a change in the dominant morphology 
of galaxy populations \citep{2015MNRAS.447.1610M} and also a transition 
between hot and cold mode accretion seen in cosmological 
hydrodynamical simulations \citep{2009MNRAS.395..160K}. 

\begin{table}
\begin{center}
\begin{tabular}{|l|c|c|}
\hline
region & $\Omega_{HI} [10^{-4} h_{70}^{-1}]$ & $\Omega_{HI}/\Omega_{HI}^{total}$ \\
\hline
total & 4.894 $\pm$ 0.469 &  1.00 \\
\hline
luminous blue &  2.543 $\pm$ 0.298  (1.099 $\pm$ 0.115) &  0.520 (0.224) \\ 
\hline
faint blue  &  1.014 $\pm$ 0.455 (1.604 $\pm$ 0.196) &  0.207 (0.328) \\ 
\hline
luminous red  &  0.764 $\pm$ 0.124  (0.653 $\pm$ 0.110) &  0.156 (0.133) \\ 
\hline
luminous bluer & 0.215 $\pm$ 0.135  (0.333 $\pm$ 0.165) &  0.044 (0.068) \\ 
\hline
faint bluer &  0.167 $\pm$ 0.126 (0.957 $\pm$ 0.233) &   0.034 (0.196) \\  
\hline
faint red   &  0.094 $\pm$ 0.050 (0.144 $\pm$ 0.038) &   0.019 (0.029) \\
\hline
dark        &  0.162 $\pm$ 0.137 &  0.033 \\  
\hline
\end{tabular}
\end{center}
\caption{The contribution of different populations to $\Omega_{\text{HI}}$. Column 2 
is the estimate of $\Omega_{\text{HI}}$ from a single population and column 3 is the 
fractional contribution to $\Omega_{\text{HI}}^{\text{tot}}$. The estimated values for
the $1\sigma$ sample is in brackets.}
\label{tab_omegah1}
\end{table}

One interesting result that we have quantified in the last section is the 
non-negligible HI content of red galaxies. The red galaxies dominate 
the HIMF at the high mass end $M_{\text{HI}} \geq 10.4$ and $\sim 17\%$ 
of the HI content, $\Omega_{\text{HI}}$, is locked up in them.  
Using the HOD framework for HI, \cite{2018MNRAS.479.1627P} also 
find that the red population is the dominant population
at higher masses. Since the ALFALFA sample is an HI selected sample,
with the majority of the galaxies belonging to the blue cloud, one may ask:
why do the rarest, gas rich galaxies, predominantly belong to the red cloud?
Looking at the morphology of these gas rich red galaxies we find that these are 
predominantly spirals and lenticular  galaxies, but there also exist 
some elliptical galaxies.
A number of spirals have prominent bulges which would classify them as early-type spirals
and there are a number of galaxies which harbor dust lanes visible on their disk plane. 
A significant number of galaxies are edge-on or somewhere in between edge-on and face-on.
Indeed HI has been detected in early type galaxies  
\citep{2006MNRAS.371..157M,2007A&A...465..787O,2009A&A...498..407G,2012MNRAS.422.1835S}, 
but these do not go beyond $M_{\text{HI}} \sim 10$. \cite{2010MNRAS.408..919S} find that 
the 47\% of the total local 
SFR density is found for $M_{\text{star}} > 10$ in the GASS sample.
 CO (a tracer for $H_2$ and a proxy for star formation) 
detections have also been reported for a fraction of 
the GASS sample \citep{2011MNRAS.415...32S}.
Given that there is little correlation between (\emph{u-r}) color and $M_{\text{star}}$ 
at these masses and the fact that GASS (and ALFALFA) 
detections are predominantly in the blue cloud
\citep{2010MNRAS.403..683C} we would expect these red gas rich galaxies 
to contribute a negligible fraction to the local SFR density. This does not mean
that individually all luminous red, gas rich galaxies have low star formation rates 
but, rather, their numbers are so small that their total contribution is negligible.  
A fraction of these red galaxies would then be the ones 
transitioning from a blue star-forming phase to a red passive phase with 
little star formation and another fraction will be dusty star forming galaxies, while 
the rest would be passively evolving. The amount of reddening
would be enhanced if they are edge-on  and would redden the color 
of disky galaxies which are either transitioning to the red phase or 
contain considerable dust on their disks \citep{2008MNRAS.388.1708G,2011A&A...529A..53T}.

Although the luminous red galaxies are the dominant population at the 
high mass end ($M_{\text{HI}} \geq 10.4$) the observed counts (60\% of total) 
are only 50\% more than those of the luminous blue (40\% of total) population. When plotted 
on a logarithmic scale the differences between these two mass functions are not very large
(see top row of figure~\ref{fig_himf}). If the inclination and reddening are important
and the magnitudes are corrected for them, then a fraction of gas rich red galaxies 
would move to the luminous blue population and would bring the mass functions of these 
two populations closer to each other at the high mass end. 
Conversely if the reddening is increased due to 
inclination we would expect that the effects would be more dramatic in 
HI velocity width function. This is indeed seen 
for ALFALFA galaxies \citep{2014MNRAS.444.3559M}, where the HI velocity width 
function for the red and blue are well segregated at the high velocity end. 
We therefore believe that reddening due to dust and inclination can partially  explain 
why the red sample is the dominant population 
at the high mass end of the HIMF.

The results of section~\ref{sec_results} 
are essentially conditional HIMF integrated over a range 
in color and magnitude, which we have called the HIMF for different populations. We can 
repeat the exercise and compute the HIMF in finer intervals of color and 
magnitude to obtain 
a conditional HIMF (conditioned on luminosity and color). This will then tell us 
about the distribution of HI in the color-magnitude plane. As an application one  
can then make better estimates of the corrections applied to $\Omega_{\text{HI}}$
with the stacking methods 
\citep{2013MNRAS.435.2693R,2016MNRAS.460.2675R,2018MNRAS.473.1879R} at higher redshifts. 
A second application would be to inform a proposed HI survey, which 
galaxy populations to stack on to make a tentative detection. 

In the survey volume considered in this work we 
find that only 11\% of the red population in SDSS have HI detections
in ALFALFA. This number is 38\% for blue galaxies. On the other hand 98\% of ALFALFA 
galaxies have optical counterparts. The detections are due to a combination 
of total HI signal and observed HI velocity widths. The question we wish to ask 
is are the non-detections in the red cloud due to insufficient HI gas or due to large  
velocity widths or both? We argue that in the luminous red sample the non-detections
are due to insufficient HI gas as well as large widths and HI masses should decrease with 
either increasing stellar mass or halo mass. 
Although figure~\ref{fig_mstar-mh1}
suggests that the most massive galaxies (large stellar mass, $M_{\text{star}}$) are also 
the richest in terms of their gas content, 
this relation is biased since it is from an HI selected sample. The appropriate 
sample is the GASS sample which is selected on stellar mass.  We look at the 
$M_{\text{HI}}-M_{\text{star}}$ relation in the final data release of GASS (DR5), which is
summarized on table 1 of  \cite{2013MNRAS.436...34C}. The relation has nearly a flat slope
for $M_{\text{star}} \in [10.76,11.30]$, being slightly positive if all non-detections 
have been assigned the limiting HI mass and slightly negative if all non-detections 
are assigned zero HI mass. The last bin however has only a handful of objects
which are dominated by non-detections. However  the GASS and ALFALFA catalogs 
are relatively shallow as compared to optical surveys like the SDSS and 
would miss a significant number of massive galaxies. The tail of the stellar mass function
is dominated by the red galaxies  and at 
$M_{\text{star}} = 11.3$ the number of red galaxies is $\sim 10\times$ the blue galaxies 
\citep{2012MNRAS.421..621B}. These galaxies are probably central red galaxies 
\citep{2009ApJ...707.1595D} and would be in halos of $\log_{10} (M_{\text{halo}}/\msun) 
\simeq 14 - 14.5$ \citep{2010ApJ...717..379B} with virial temperatures 
$T_{\text{vir}} \sim 10^7 \text{K}$ and circular velocities $V_{\text{circ}} \sim 6-9 \times 10^2 \text{km.s}^{-1}$. Most of these galaxies will then be the central galaxies of 
large groups of galaxies or clusters of galaxies. Given the large virial temperatures
and large circular velocities it would be very unlikely to detect a considerable amount 
of neutral hydrogen in these systems. It is then very likely that 
the $M_{\text{HI}}-M_{\text{star}}$ relation will not asymptote to a constant as indicated
in  \cite{2013MNRAS.436...34C} but rather decrease with increasing stellar mass. 
This is suggested in the results of \cite{2017MNRAS.465..111K} and 
\cite{2019arXiv190902242S}.

If the  average 
$M_{\text{HI}}-M_{\text{star}}$ becomes a non-monotonic function of stellar mass and 
therefore halo mass, 
HI abundance matching techniques, used to obtain $M_{\text{HI}}-M_{\text{halo}}$ relation
 \citep{2011MNRAS.415.2580K,2017MNRAS.470..340P}, will break down. The HI HOD models 
which also assume a step-like function (with the help of the error function)  
\citep{2017ApJ...846...61G,2018MNRAS.479.1627P} for the average occupation 
of centrals, may need to be revised. A log normal form  for the mean  occupation 
function for centrals was compared to the step-like parameterization in the context
of describing quasar clustering \citep{2013ApJ...778...98S}, but it was found that the HOD 
parameters were not well constrained. Only more direct observations will shed light 
on the HI content of these massive galaxies and hopefully provide better inputs 
for the HOD parameterization.

We end this section by discussing how the sensitivity limits of ALFALFA may  
affect our results. We start with targeted HI observations, more
sensitive than ALFALFA, that look at the HI content 
of massive galaxies and also luminous early type galaxies (ETGs) in the local Universe.
In the GASS survey \citep{2010MNRAS.403..683C, 2013MNRAS.436...34C} 
the targets were selected by  stellar mass, $10 < M_{\text{star}} < 11.5$ 
from an area common to 
SDSS \citep{2009ApJS..182..543A}, \emph{GALEX} \citep{2005ApJ...619L...1M} and 
ALFALFA \citep{2005AJ....130.2598G} in the redshift range $ 0.025 < z < 0.05$.
The detection limit for GASS was set to a very low HI gas mass fraction of 
$f_{\text{HI, lim}}=M_{\text{HI, lim}}/M_{\text{star}}>0.015$ for $M_{\text{star}}>10.5$,
 and a constant HI gas mass limit of $M_{\text{HI, lim}}=8.7$ 
for smaller stellar mass targets, 
which translates roughly to upper HI mass limits $8.7\leq M_{\text{HI, lim}}\leq 9.7$ 
for the non-detections. 
Since the maximum limiting mass for the more sensitive 
GASS survey, $M_{\text{HI, lim}}^{\text{max}} = 9.7$, is below the
characteristic mass, $M_*$, for the populations considered in this work  
(see table~\ref{tab_schecterfit}), the sensitivity limit of 
ALFALFA does not affect the large mass ($ M_{\text{HI}} > M_* $) end of the HIMF. 

As of the final data release from the total of 666 targeted galaxies 
in GASS, 287 are non detections \citep{2013MNRAS.436...34C}. The non-detections 
span the entire targeted stellar mass range, 
they are mostly redder in color (NUV-r), have larger concentration index and the 
detection fraction is about 70\% for $M_{\text{star}} < 10.7 $ and drops to 40\% beyond that.
As discussed earlier the detection fraction in ALFALFA is 
 11\%(38\%) for red(blue) galaxies in SDSS. Although lower than GASS, 
this is consistent with the trend seen in GASS where the 
non-detections are dominated by red galaxies. We would then expect that there 
is a non-negligible population of massive galaxies $M_{\text{star}} > 10$, dominated 
by the bright red population, which could host HI gas masses upto 
$M_{\text{HI, lim}} = 9.7$, and have 
not been detected by ALFALFA due to their large velocity widths. The HI mass of these 
objects would then be anywhere in between 0 and $M_{\text{HI, lim}}$. 

The $\text{ATLAS}^{\text{3D}}$ HI survey \citep{2012MNRAS.422.1835S}  complements the 
results of the GASS survey by reporting HI masses of ETGs
(elliptical E and lenticular S0). 
The  $\text{ATLAS}^{\text{3D}}$ ETG sample is morphologically 
selected from a volume limited parent sample of 871 nearby 
($D < 42\text{Mpc}, \left| \delta - 29^{\circ} \right| < 35^{\circ}, 
\left| \text{b} \right| > 15^{\circ}$) galaxies brighter than
$M_{\text{K}} < -21.5$ which translates to stellar masses $M_{\text{star}} \geq 9.78$ 
\citep{2011MNRAS.413..813C}.
Of the parent sample of 871 galaxies, 611 (70\%) are spirals, 192 (22\%) 
are lenticular and 68 (8\%) are elliptical. 95\% of the ETGs lie on the red sequence
(as demarcated by eq.~\ref{eq_tanh_cut}).
The $\text{ATLAS}^{\text{3D}}$ HI
observations  are done for 166 ETGs ($\delta > 10^{\circ}$) with the 
Westerbork Synthesis Radio Telescope of which there are 53 (32\%) detections and 
113 (68\%) non detections. The $\text{ATLAS}^{\text{3D}}$ sample therefore represents
massive, red, E or S0 type galaxies in the local Universe. The HI detections 
of ETGs have a broad distribution in the range $M_{\text{HI}} \in [7,9.5]$.
On the other hand the HI mass 
(accessed from the HyperLeda\footnote{http://leda.univ-lyon1.fr/} database) distribution 
of  spirals in the parent sample are much narrower, having a peak at 
 $M_{\text{HI}} \sim 9.3$  and a tail at $M_{\text{HI}} \sim 8$ \citep{2012MNRAS.422.1835S}.
The distribution of limiting masses of HI non-detections is in the range   
$M_{\text{HI, lim}} \in [6.5,8.5]$ peaking at $M_{\text{HI, lim}} \sim 7.1$. 
The HI distributions of detected ETGs and spirals overlap significantly which means
that a non-negligible fraction of ETGs contain as much HI as in spirals. The HIMF of ETGs
have a relatively low value of the characteristic mass $M_* = 9.26$, which is a 
factor of 5 smaller than $M_*=9.96$ for the full ALFALFA population and 
a factor of 6 smaller than $M_*=10.04$ for the luminous red population of ALFALFA.
The difference between HI column density distribution ($N_{\text{HI}}$) 
for ETGs and spirals in the  $\text{ATLAS}^{\text{3D}}$
survey is significant. The characteristic column density (when the $N_{\text{HI}}$ 
distribution 
is parametrized by a Schecter function) is $N* = 9.2 \times 10^{19}\text{cm}^{-2}$ for ETGs 
and $N* = 1.03 \times 10^{21}\text{cm}^{-2}$. Therefore the HI in 
gas rich ETGs is rarely as dense as the average column densities of 
spirals \citep{2012MNRAS.422.1835S}. Given that the 
GASS and the  $\text{ATLAS}^{\text{3D}}$ HI 
surveys are targeted surveys, more sensitive than ALFALFA, and are specifically looking
at the HI gas content of massive galaxies we conclude that it is very unlikely
that ALFALFA has missed out any galaxy with masses $M_{\text{HI}} > 10$ due to an unusually 
large velocity width. However due to their moderate amounts of HI 
individual detections are less frequent in ALFALFA as compared to these surveys.

Finally it is worth considering how a different choice of the sensitivity limit
\cite[e.g. dictated by high S/N objects Code 1 objects in 
this work and in ][]{2010ApJ...723.1359M} 
of ALFALFA in turn affects the HIMF. 
For example, could a lower sensitivity 
limit, by including lower S/N Code 2 objects, alter the HIMF? 
\cite{2011AJ....142..170H} 
addressed this question by considering Code 1 and 2 sources, which in turn  
lowers the sensitivity limit compared to Code 1 only objects 
\citep[see eqs. 6 and 7 and discussion in section 6 of][]{2011AJ....142..170H}.
Both the amplitude and characteristic mass of HIMF remained unchanged, however 
the faint-end slope of the HIMF decreased from $\alpha = -1.33 \pm 0.02$ (Code 1)
to $\alpha = -1.29 \pm 0.02$ (Code 1 and 2). They concluded that there is little 
value in adding lower S/N objects in the analysis and that statistical estimators
like the 2DSWML method are robust to such changes. 

In a separate analysis \cite{2012ApJ...759..138P} explored if there were systematic 
differences in estimating the HIMF by considering an HI-selected sample and 
an optically-selected sample. The HI-selected sample is the $\alpha.40$ sample
and the corresponding estimate of the HIMF is found with methods similar to the one
described in this work. The optically-selected sample consists of all the HI detections
and non-detections. The non-detections are the SDSS galaxies in the same volume as ALFALFA 
which do not have HI detections. These galaxies are assigned a lower and upper limiting
HI mass ($M_{\text{HI, lim}}^{\text{min}},M_{\text{HI, lim}}^{\text{max}}$). The lower limit is 
$M_{\text{HI, lim}}^{\text{min}} = 0$. The upper limit is computed by converting the detection 
limit, which is the 25\% completeness limit \cite[eqs. 5 and 7 in][]{2011AJ....142..170H}, 
to an HI mass. 
To estimate  $M_{\text{HI, lim}}^{\text{max}}$ we need  a distance
(which exists) and velocity width, $W_{50}$ (which has to be estimated).
\cite{2012ApJ...759..138P} used the average stellar mass Tully-Fisher relation, 
$M_{\text{star}}-V_{rot}$ for the $\alpha.40$ galaxies to estimate $W_{50}$ for the 
non-detections, after accounting for inclination effects. 
The optically-selected sample then consists of ALFALFA detections
and two estimates of HI masses for the non-detections. The optically-selected sample
is an r-band flux limited sample which has a different selection function as compared
to the HI-selected sample. The HIMF estimated from the optically-selected and HI-selected
sample should broadly be consistent with each other. 
A naive expectation is that the HIMF from 
the HI selected-sample should lie in between the two estimates of the HIMF from 
the optically-selected sample. In the limit that the assigned 
lower and upper limiting HI masses 
approach the {\emph{true}} HI mass of the undetected source we expect 
the HIMF estimated from the optically-selected sample to approach that 
of the HI-selected sample.
We also point out at the high mass end, all the estimates of the 
HIMF should be the same. This is demonstrated in figure 6 of \cite{2012ApJ...759..138P}.
Given the uncertainties associated in obtaining $M_{\text{HI, lim}}^{\text{max}}$ which 
use average scaling relations and inclination effects (which are prone to errors) 
we do not comment further on the differences between the HIMF from the two samples. 
Based on the arguments presented above, 
we believe that although the sensitivity of ALFALFA affects the individual 
detections as compared to more sensitive targeted surveys, the estimates 
of the HIMF for  different populations considered here are robust.

\section{Summary}
\label{sec_summary}
In this work we have measured the HIMF of different galaxy populations picked from  
the color-magnitude  plane. The galaxies considered were from a local volume common to 
ALFALFA and the SDSS surveys in the redshift range $z \in [0,0.05]$. After putting the 
relevant cuts in quality of detection, volume and completeness the 
final sample analyzed consists of 7857 galaxies. We divided the total sample 
first into luminous and faint populations (cut in magnitude) 
and these were further split into three colors: red, 
blue and bluer. This forms a disjoint set of six populations. A seventh population which 
we call a dark population is one which does not have any optical counterparts in SDSS 
but has a detection in ALFALFA. The union of these seven populations is the total sample 
of 7857 galaxies.
We have further considered a second sample which redefines the six populations 
by shifting the boundaries defining them. We have called them the $1\sigma$ and $1.5\sigma$
samples (section~\ref{subsec_subsamples}). The reason for doing this is to illustrate
that our sample definition does not change the qualitative results that we report. 
The reason for splitting the full blue cloud into four (faint/luminous for blue/bluer)
instead of just two (faint/luminous for all blue) 
was because we wanted to see the systematic effect 
of the tail of the blue population (especially the faint end) on our results. We indeed 
find that based on how we define our sample the faint bluer population becomes the dominant
population at the low mass end of the HIMF (figure~\ref{fig_himf} and 
tables~\ref{tab_schecterfit}\&\ref{tab_omegah1}). 

We summarize our results below:

\begin{itemize}
\item For a fixed range in 
  color, in the color-magnitude plane 
  (e.g. red, blue or bluer), the HIMF of the luminous population 
  dominates over their corresponding faint counterparts at the knee and the large mass end
  whereas the faint populations dominate at the low mass end. 

\item  For a fixed magnitude interval,  in the color-magnitude plane
  (e.g. luminous or faint) 
  there is no systematic trend at the low 
  mass end and the knee of the HIMF, with decreasing color, for the faint population. 
  However for the luminous population, 
  we see that the HIMF at the high mass end increases with increasing color.
  The luminous red population is the dominant population at this end. 

\item The luminous red population dominates the total HIMF at $M_{\text{HI}} \geq 10.4$.
  When combined with the faint red sample it contributes about  $\sim 16-17\%$ of the 
  $\Omega_{\text{HI}}$ budget. 
  The dark population contributes $\sim 3\%$ to $\Omega_{\text{HI}}$.
  This has implications for detections done with stacking at higher redshifts
  which would target the blue star forming cloud for a first detection.
 
\item The total blue cloud (blue and bluer) represents about $\sim 80\%$ 
of the $\Omega_{\text{HI}}$ budget.

\item In the mass range about the knee, $M_{\text{HI}} \in [8,10.4]$, the dominant
  populations are the faint and luminous blue populations with the latter dominating
  at larger masses in this interval. Their total contribution to $\Omega_{\text{HI}}$ is 
  $\sim 55-70\%$ depending on sample definitions. 

\item The dominant populations contributing to the low mass end of the HIMF 
  are the faint blue and faint bluer populations, the latter being dominant only for 
  the $1\sigma$ sample definition. 

\item The luminous bluer and faint red populations are subdominant populations
  contributing a total of $\sim 6-10\%$ to $\Omega_{\text{HI}}$. For the 1.5$\sigma$ 
  sample the fraction of luminous bluer(faint red) 
  galaxies in ALFALFA to that in SDSS is 41\%(28\%). In comparison the detection rate 
  of ALFALFA in the red cloud is 11\% and the blue cloud is 38\%. This shows that 
  although the detection rate of ALFALFA for the faint red population is higher 
  than the luminous red population their contribution to $\Omega_{\text{HI}}$ is small.
  This is because the number density of these galaxies is small and the HI mass of these 
  galaxies is also small. On the contrary the luminous blue galaxies are HI rich 
  (figure~\ref{fig_mstar-mh1}) on average but due to their small number densities 
  (figure~\ref{fig_himf}) they too contribute little to the $\Omega_{\text{HI}}$ budget.

\end{itemize}

It would be interesting to see if simulations \citep{2017MNRAS.467..115D}, SAMs 
\citep{2017MNRAS.465..111K} and HOD models \citep{2018MNRAS.479.1627P} 
are able to reproduce our results
which provide additional constraints on the population of HI selected galaxies.
In section~\ref{sec_discussion} we argued that the effect of dust and inclination 
are responsible for the luminous red population dominating the 
total HIMF at large masses. 
This was based on the results of \cite{2013MNRAS.436...34C,2014MNRAS.444.3559M}
but needs to be confirmed with a more detailed followup.
In a forthcoming paper we will report on a more detailed analysis of 
the HI velocity width 
function. We are also working on measuring the conditional 
(conditioned on color or magnitude or both) HIMF
which will put additional constraints on the properties 
of gas bearing galaxies.

\section*{ACKNOWLEDGMENTS}
We would like to thank  R. Srianand and A. Paranjape for the many useful 
discussions that we had throughout this work. 
SD would like to thank R. Srianand for giving the hands-on training 
on obtaining and processing SDSS data. We would like to acknowledge 
discussions with J. S. Bagla and N. Kanekar. The research of NK 
is supported by the Ramanujan Fellowship awarded by the 
Department of Science and Technology, Government of India.
All the analyses were done on the xanadu server funded by the Ramanujan Fellowship.
NK acknowledges the support from the Inter University Centre for Astronomy and 
Astrophysics (IUCAA) associateship programme. We would like to thank the referee for useful
comments which helped in improving the presentation of this work.

We would like to acknowledge the work of the entire ALFALFA collaboration 
in observing, flagging, and extracting the properties of galaxies that this paper 
makes use of. This work also uses data from SDSS DR7.
Funding for the SDSS and SDSS-II has been provided by the Alfred P. Sloan Foundation, 
the Participating Institutions, the National Science Foundation, 
the U.S. Department of Energy, the National Aeronautics and Space Administration, 
the Japanese Monbukagakusho, the Max Planck Society, and the Higher 
Education Funding Council for England. The SDSS Website is 
http://www.sdss.org/. The SDSS is managed by the Astrophysical Research 
Consortium for the Participating Institutions.

\appendix
\section{The 2DSWML method to Calculate the HI Mass Function}
\label{appendix}
We now proceed in describing our implementation of the 2DSWML method
\citep{2003AJ....125.2842Z, 2010ApJ...723.1359M}.
Assuming a bivariate distribution $\phi(\mhi,\w50)$ the 
probability of detecting a galaxy '\emph{i}' with HI mass $\mhi^i$
and profile width $\w50^i$ at distance $D^i$ is

\beq
p_{i} = \frac{\phi \left(\mhi^i, \w50^i\right)}
{\int_{\w50=0}^{\infty} 
\int_{M_{\text{HI}}=M_{\text{HI},lim} \left( D^i, W_{50}^i\right)}^{\infty} 
\phi \left( M_{\text{HI}}, W_{50} \right) d M_{\text{HI}} \; dW_{50}}
\label{eq_probi}
\eeq

\begin{figure}
  \begin{tabular}{c}
    \includegraphics[width=3.2truein]{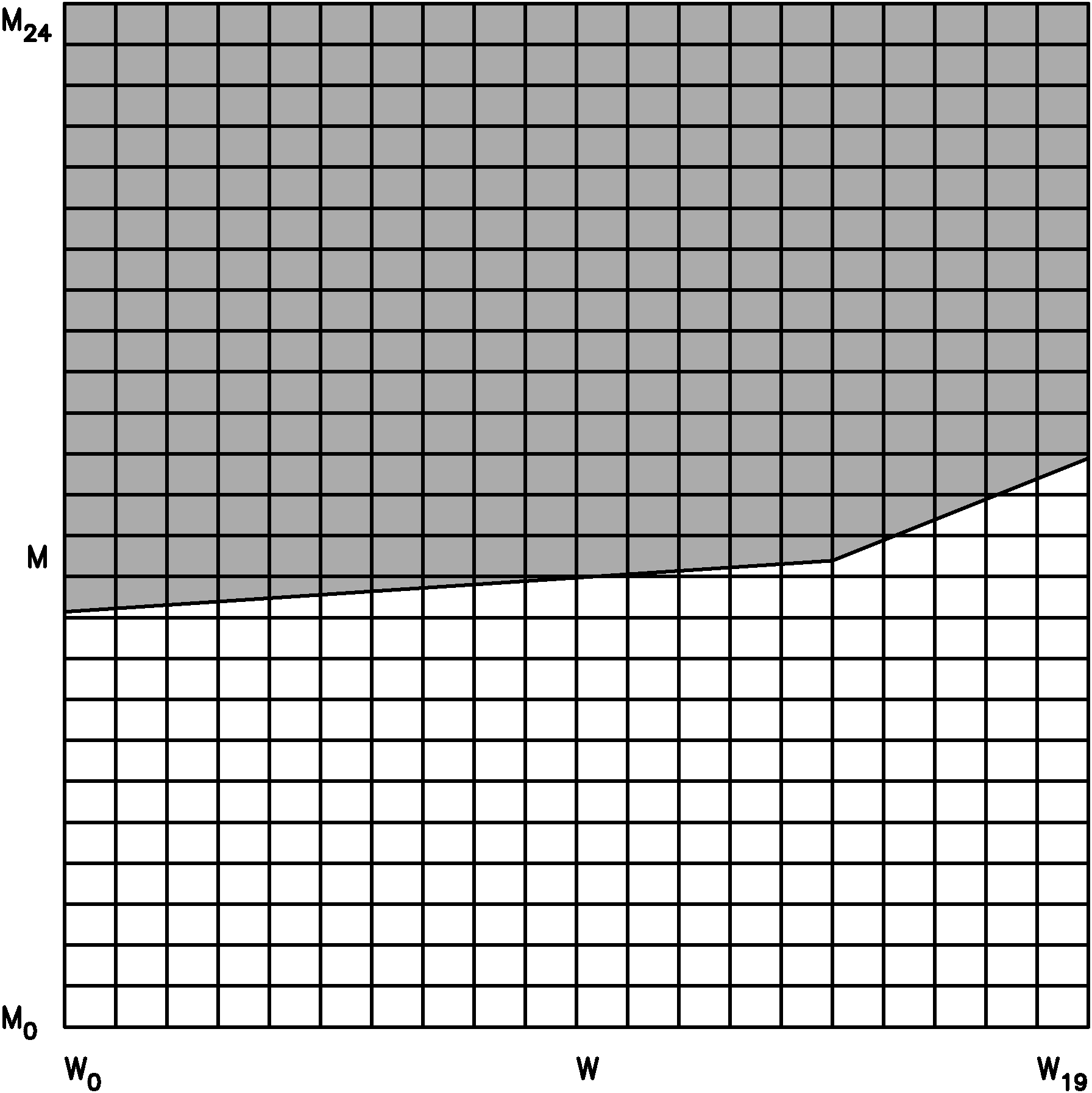}
  \end{tabular}
  \caption{The completeness relation (\emph{broken solid line}) is shown in the M-W plane.
This particular object '\emph{i}' has the following properties: $\log_{10}\left[\mhi^i/\msun\right] = $ 8.73, $\left[\w50^i/\text{km.s}^{-1}\right] = 93, \left[D/\text{Mpc} = 27\right]$. The shaded(white) region is the area accessible(inaccessible) to this object in the \emph{M-W} plane.}
  \label{fig_cmw}
\end{figure}

We now need to discretize the above equation for the 2DSWML 
method. We will begin by considering the distribution of galaxies
in bins of $M=\log_{10} \left[ \mhi/\msun\right]$
and $W=\log_{10}\left[\w50/\left(\text{km.s}^{-1}\right)\right]$.
The number of bins are $N_M$ \& $N_W$ 
and the bin widths are $\Delta M$ \& $\Delta W$ respectively.
Therefore the two dimensional distribution 
can be parameterized by $\phi_{jk}$,
where $j=0,1,2,...,N_M-1$ and $k=0,1,2,...,N_W-1$.
In this analysis we have taken 10 bins per dex in velocity width 
and 5 bins per dex in mass. Eq~\ref{eq_probi} for the 2DSWML is now
\beq
p_i = \frac{\Sigma_j \Sigma_k V_{ijk} \phi_{jk}}
{\Sigma_j \Sigma_k H_{ijk} \phi_{jk} \Delta M \Delta W}
\eeq
Here $V_{ijk}$ ensures that galaxy '\emph{i}' is only populated 
in its corresponding '\emph{j-k}' bin.  
\begin{eqnarray}
V_{ijk} = \begin{cases}
    1, & \text{if galaxy $i$ is a member of mass bin $j$} \\
       & \text{and profile width bin $k$}.\\
    0, & \text{otherwise}.
  \end{cases}
\label{eq_hijk}
\end{eqnarray}
$H_{ijk}$ is a weight corresponding to galaxy '\emph{i}'
in the '\emph{j-k}' bin and takes values from 0 to 1.
It appears so that the integral in the denominator of eq.~\ref{eq_probi}
can be done in the entire \emph{M-W} plane after convolving with the 
completeness function $C^i$ in the \emph{M-W} plane. An example  
for one of the galaxies in our sample is shown in figure~\ref{fig_cmw}. The 
solid broken line is the completeness relation for this object in the \emph{M-W} plane.
The shaded(white) area is the area accessible(inaccessible) to this object. 
The shaded area is given by: 
\baq
&&\Delta M\Delta W \sum_{k=0}^{N_W-1}\sum_{j=0}^{N_M-1} H_{ijk} = \nonumber \\
&&\int_{W=W_0}^{W=W_{N_W-1}} \int_{M=M_0}^{M=M_{N_M-1}} C^i(M,W)\text{d}M\text{d}W 
\eaq

\begin{table}[h]
\begin{center}
\begin{tabular}{|l|c|c|}
\hline
   No. & Conditions & $H_{ijk}$ \\
\hline
   C1 &	\parbox[c]{10em}{\includegraphics[width=0.15\textwidth]{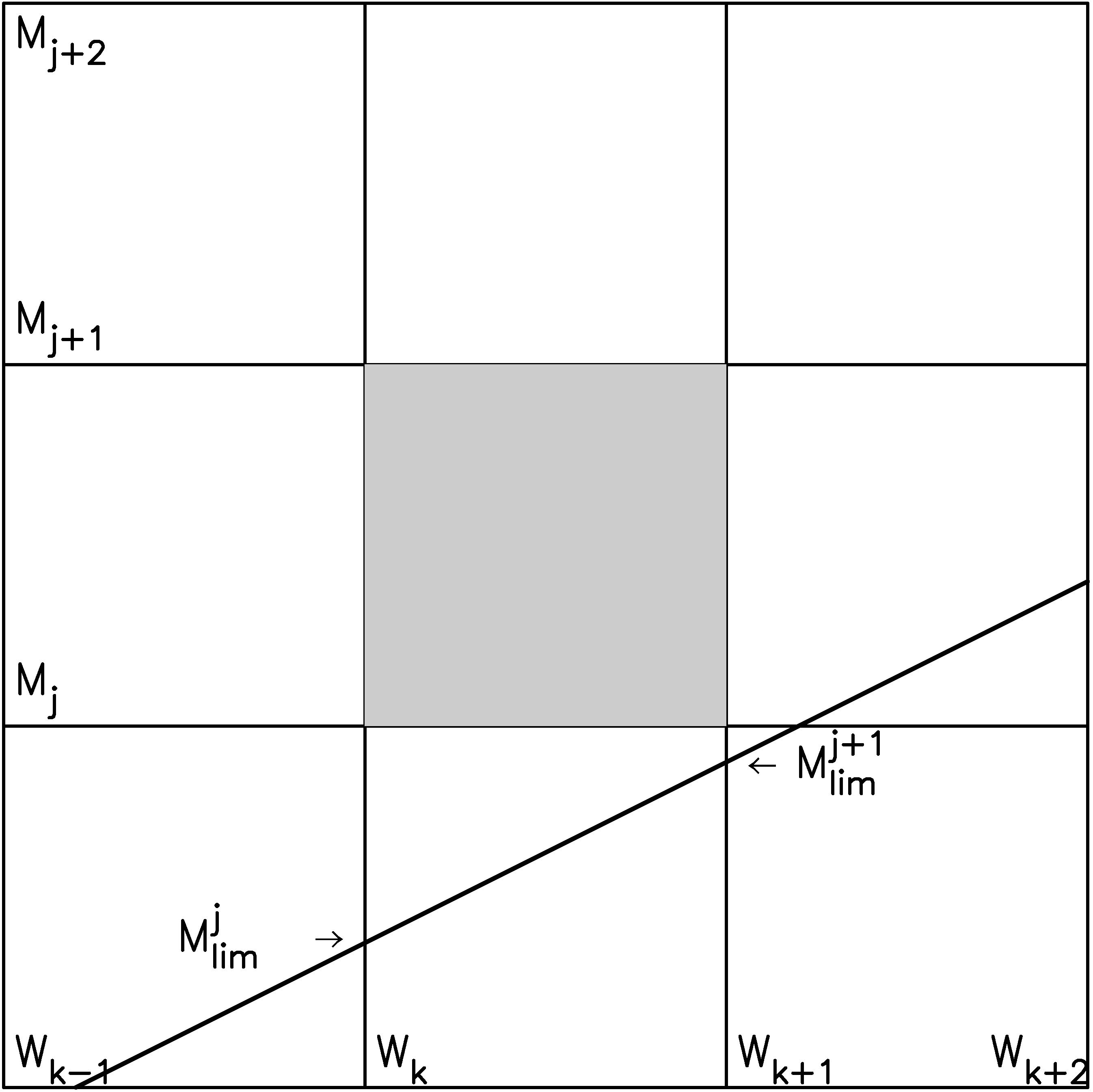}} & $1.0$ \\
\hline
   C2 &	\parbox[c]{10em}{\includegraphics[width=0.15\textwidth]{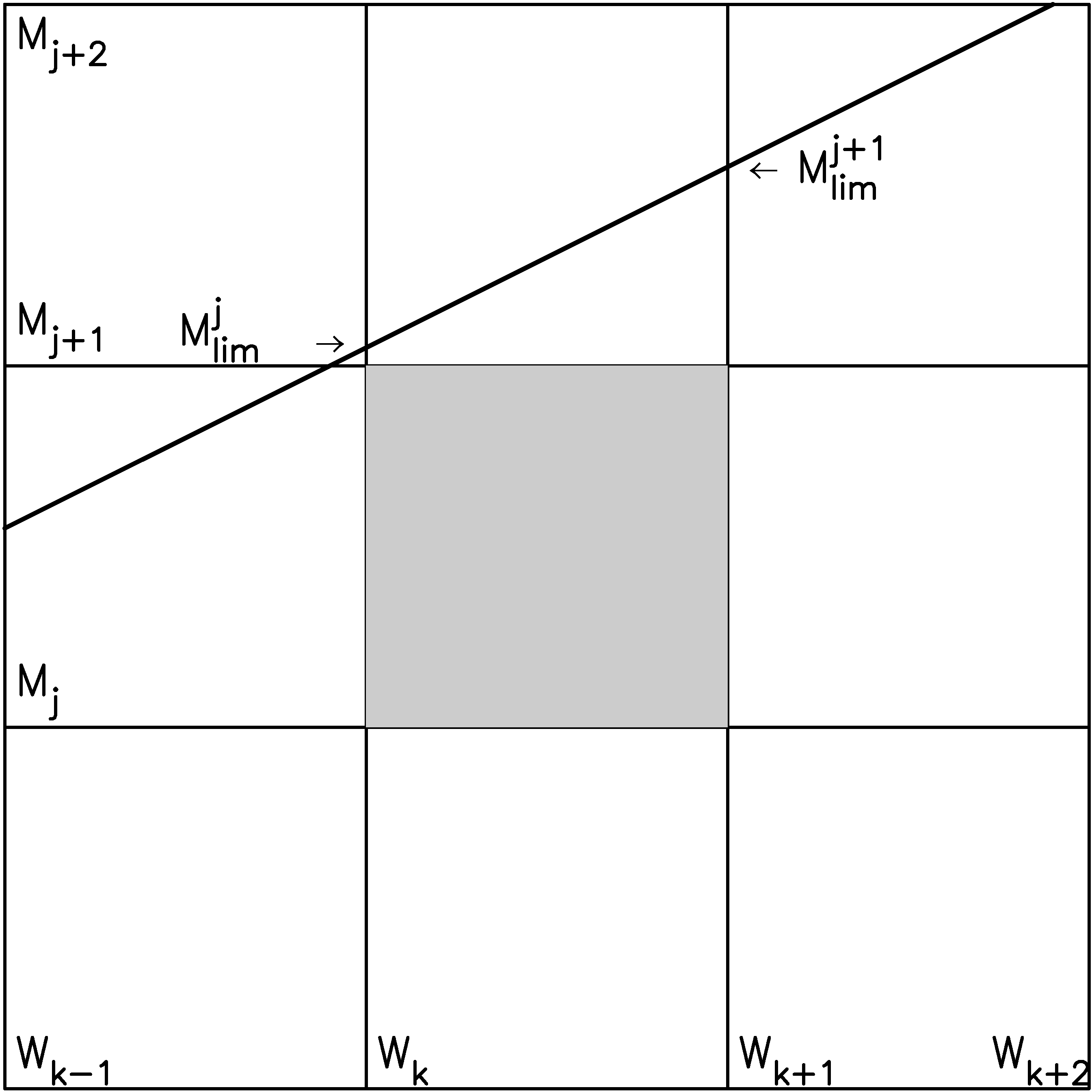}} & $0.0$\\
\hline
   C3 &	\parbox[c]{10em}{\includegraphics[width=0.15\textwidth]{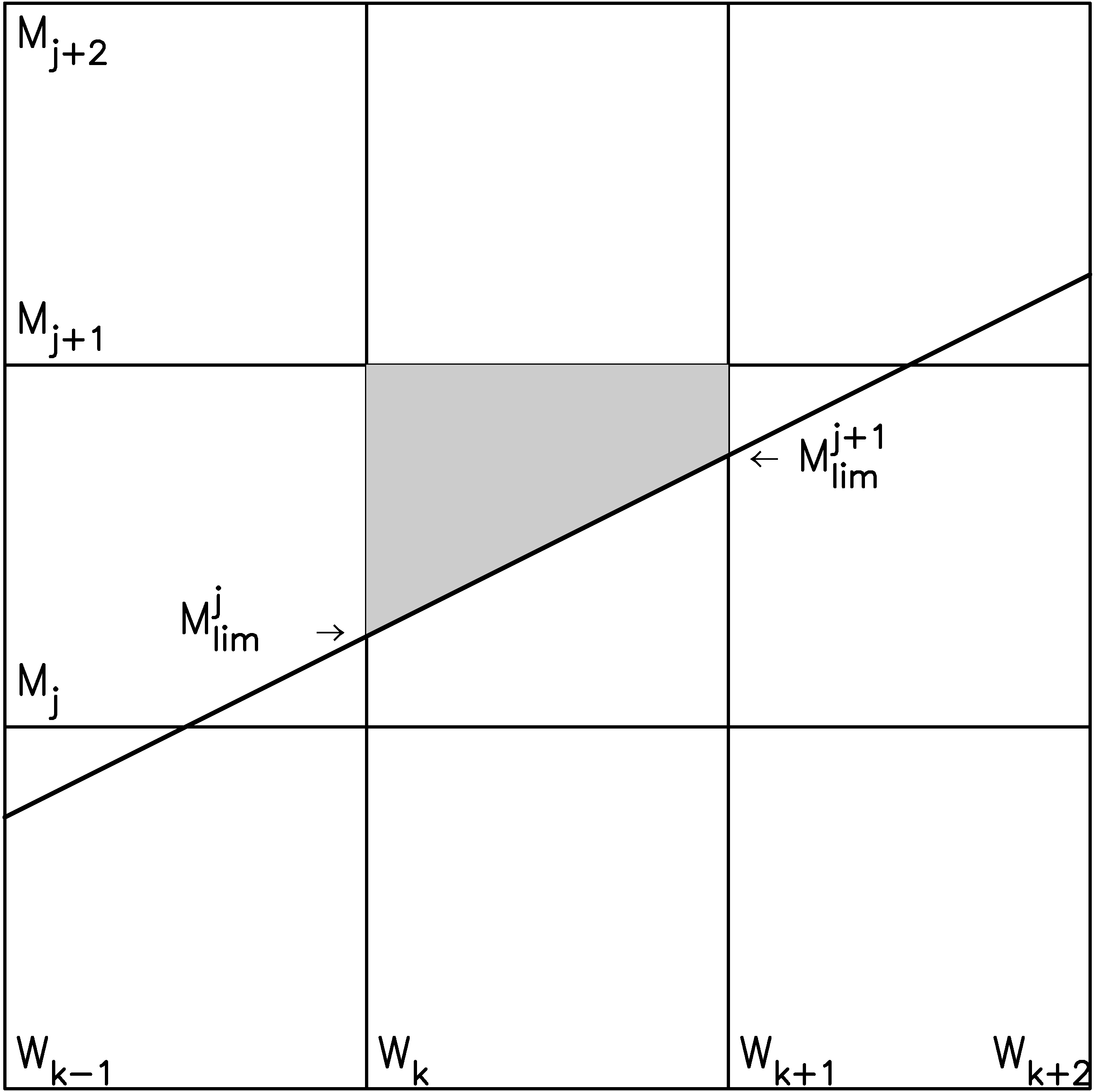}} & $\frac{2 M_{j+1} - M_{lim}^{j+1} - M_{lim}^j}{2 \Delta M}$\\
\hline
   C4&	\parbox[c]{10em}{\includegraphics[width=0.15\textwidth]{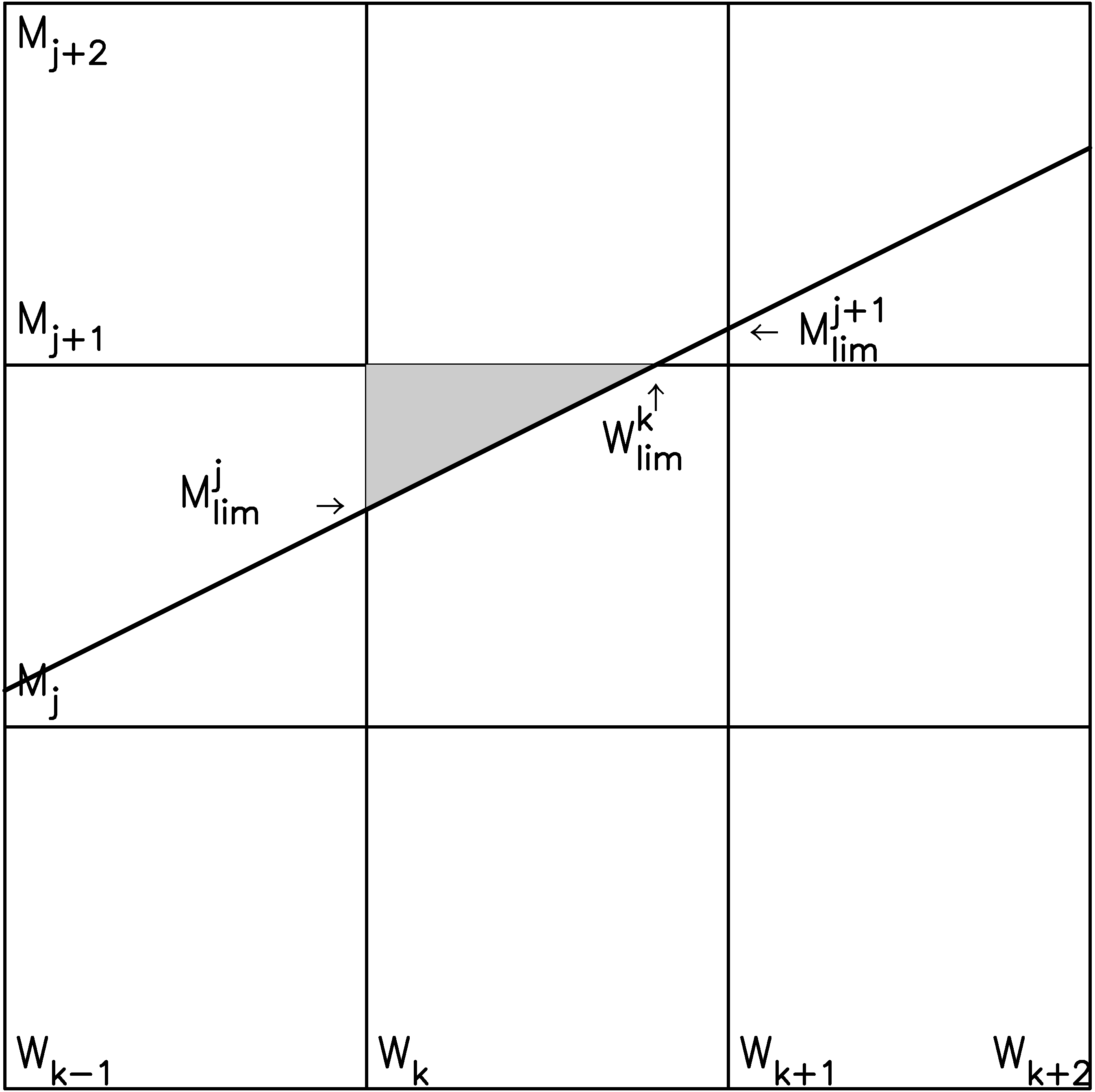}} & $\frac{\left( M_{j+1} - M_{lim}^j \right)\left( W_{lim}^{k+1} - W_k\right)}{2 \Delta M \Delta W}$\\
\hline
   C5&	\parbox[c]{10em}{\includegraphics[width=0.15\textwidth]{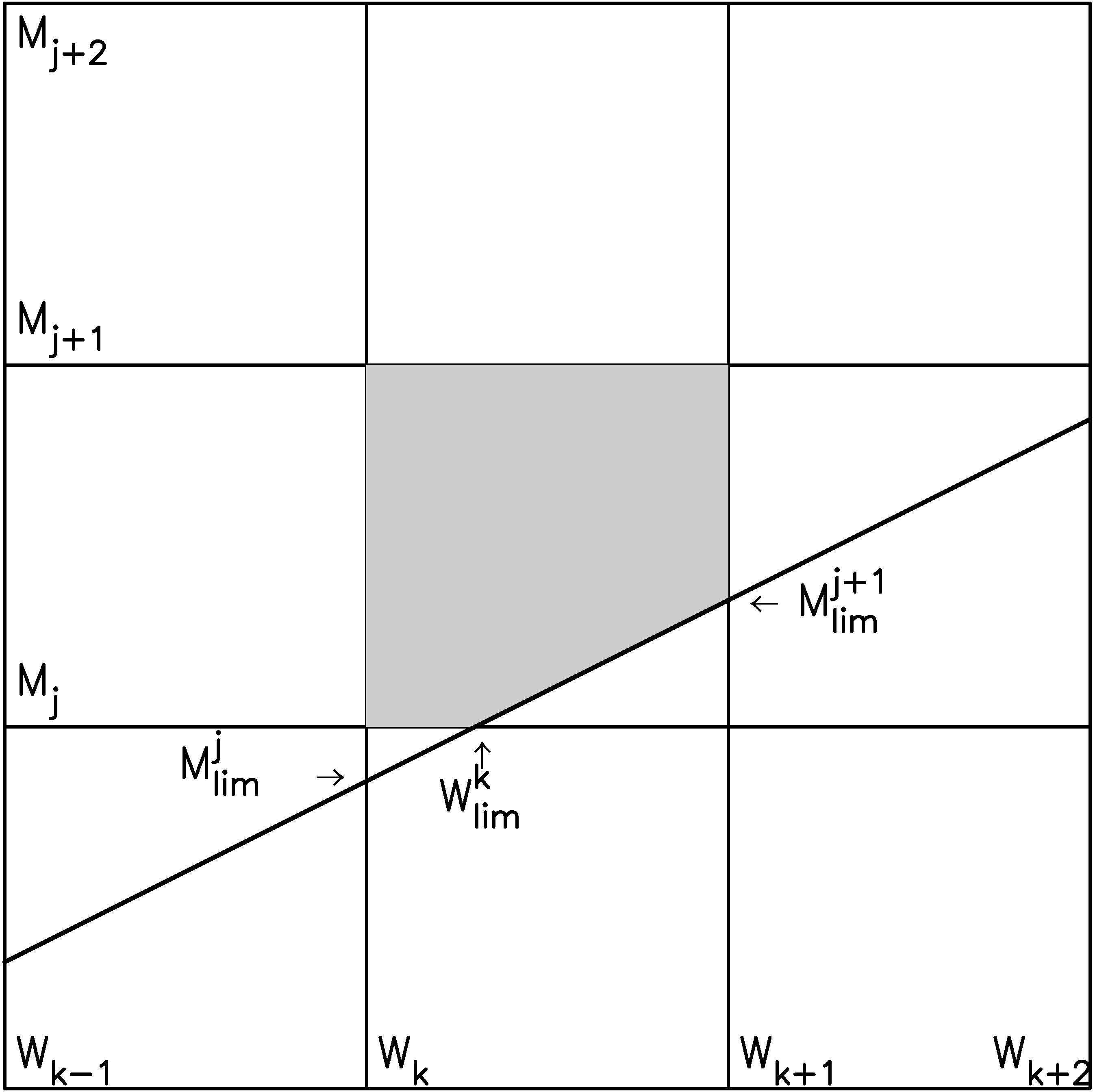}} & $1.0 - \frac{\left( M_{lim}^{j+1} - M_j\right) \left(W_{k+1} - W_{lim}^k\right)}{2 \Delta M \Delta W}$\\
\hline
   C6 &	\parbox[c]{10em}{\includegraphics[width=0.15\textwidth]{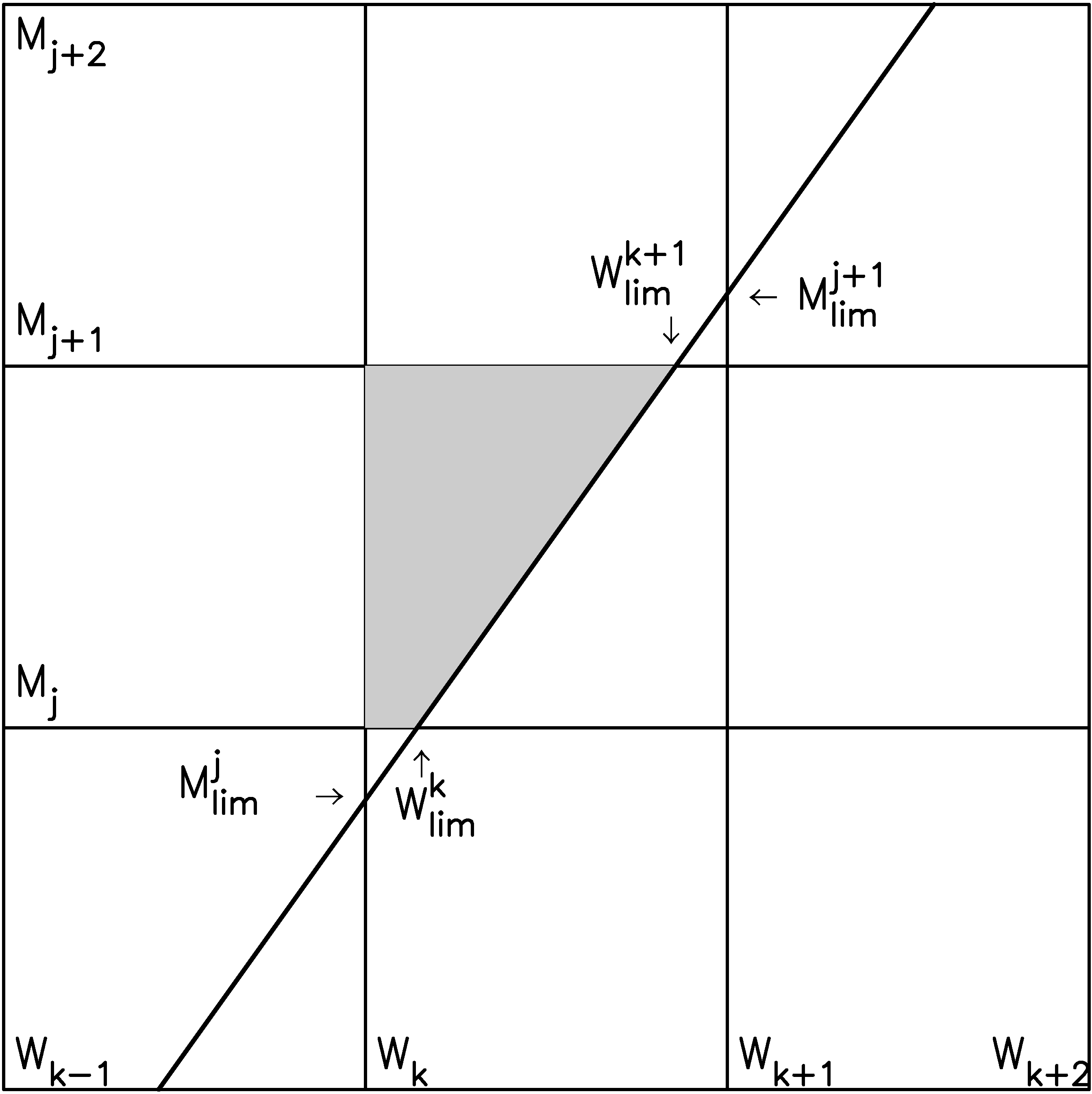}} & $\frac{W_{lim}^k-W_k}{\Delta W} + \frac{W_{lim}^{k+1} - W_{lim}^k}{2 \Delta W}$\\
\hline
\end{tabular}
\end{center}
\caption{Values of $H_{ijk}$ (column 3) are shown diagrammatically (column 2).The shaded 
region corresponds to  the \emph{j-k} bin of interest. Cases 1 and 2 (first two rows) 
take values 1 and 0,  the completeness curve (line) lies below or above the square and never intersects it. 
Cases 3-6 give fractional values of $H_{ijk}$ since the \emph{j-k} bin of interest 
intersects the completeness curve. The points of intersection are denoted 
by $W_{\text{lim}}^j, M_{\text{lim}}^k$ if it does not exactly intersect 
on the bin edges $W_k, M_j$.
}
\label{tab_hijk}
\end{table}

Computing $H_{ijk}$ is a straightforward exercise and we have shown it as diagrams
in table~\ref{tab_hijk}. The shaded area (column 2) in each case 
is the \emph{j-k} bin of interest
for which $H_{ijk}$ has to be computed. In cases 1 and 2, the values of $H_{ijk}$ 
are either 1 or 0. This is because the completeness line does not intersect the square
and lies below or above that particular bin of interest. In cases 3-5 the 
square of interest intersects with the completeness curve and the shaded area is 
the value that $H_{ijk}$ takes (column 3) which are fractional values. 
The points of intersection are denoted 
 as $W_{\text{lim}}^k, M_{\text{lim}}^j$ if it does not exactly intersect 
on the bin edges $W_k, M_j$. Case 6 is introduced and assumes a completeness slope greater 
than 1, which is not the case for ALFALFA. We also note that the completeness relation
eq.~\ref{eq_completeness} has a change in slope at  $\log_{10}\left[\w50/\left(\text{km.s}^{-1}\right)\right] = 2.5$, which coincides with the edge of the bin in $\w50$ for our choice.
Table~\ref{tab_hijk} assumes this so that no further cases are considered.

In the 2DSWML one wishes to find $\phi_{jk}$ that maximizes the joint probability
or likelihood of finding all the galaxies in the sample simultaneously 
\begin{eqnarray} \label{eq_likelihood}
\mathcal{L} &=& \prod_{i=1}^{N_g} p_i
\end{eqnarray}
Using eq.~\ref{eq_probi} the joint likelihood is 
\begin{eqnarray} \label{eq_jointlikelihood}
\mathcal{L} &=& \prod_{i=1}^{N_g} 
\frac{\Sigma_j \Sigma_k V_{ijk} \phi_{jk}}
{\Sigma_j \Sigma_k H_{ijk} \phi_{jk} \Delta m \Delta w}
\end{eqnarray}
To obtain $\phi_{jk}$, 
we maximize rather the log-likelihood 
\begin{eqnarray} \label{eq_loglikelihood}
\ln \mathcal{L} = \sum_{i=1}^{N_g} 
\ln \left( \frac{\Sigma_j \Sigma_k V_{ijk} \phi_{jk}}
{\Sigma_j \Sigma_k H_{ijk} \phi_{jk} \Delta m \Delta w} \right)
\end{eqnarray}
This gives us
\begin{eqnarray} \label{eq_phijk}
\phi_{jk} &=& \left[\Sigma_i V_{ijk}\right] 
\left[\Sigma_i \frac{H_{ijk}}
{\Sigma_m \Sigma_n H_{imn} \phi_{mn}}\right]^{-1} \nonumber\\
&=& n_{jk}\left[\Sigma_i \frac{H_{ijk}}
{\Sigma_m \Sigma_n H_{imn} \phi_{mn}}\right]^{-1}
\end{eqnarray}
where, $n_{jk} = \sum_i V_{ijk}$ is the number of galaxies in 
mass bin $j$ and profile width bin $k$. $\phi_{jk}$ are iteratively
determined from eq.~\ref{eq_phijk}. To start the iteration
we set the initial value of on the RHS of eq.~\ref{eq_phijk} to be
$\phi_{mn}^{in} = n_{jk}/\left[V_{\text{surv}}\Delta M \Delta W\right]$.
We set a relative tolerance for convergence of $10^{-3}$ for all $\phi_{jk}$.
We find that $\phi_{jk}$ converges within 20 iterations.
Finally the HIMF is obtained by summing over the velocity width bins \emph{k}.
\begin{eqnarray}\label{eq_phij}
\phi_j &=& \sum_k \phi_{jk} \Delta w
\end{eqnarray}

\subsection{Normalization of HIMF}
\begin{figure}
\centering
\includegraphics[width=\columnwidth]{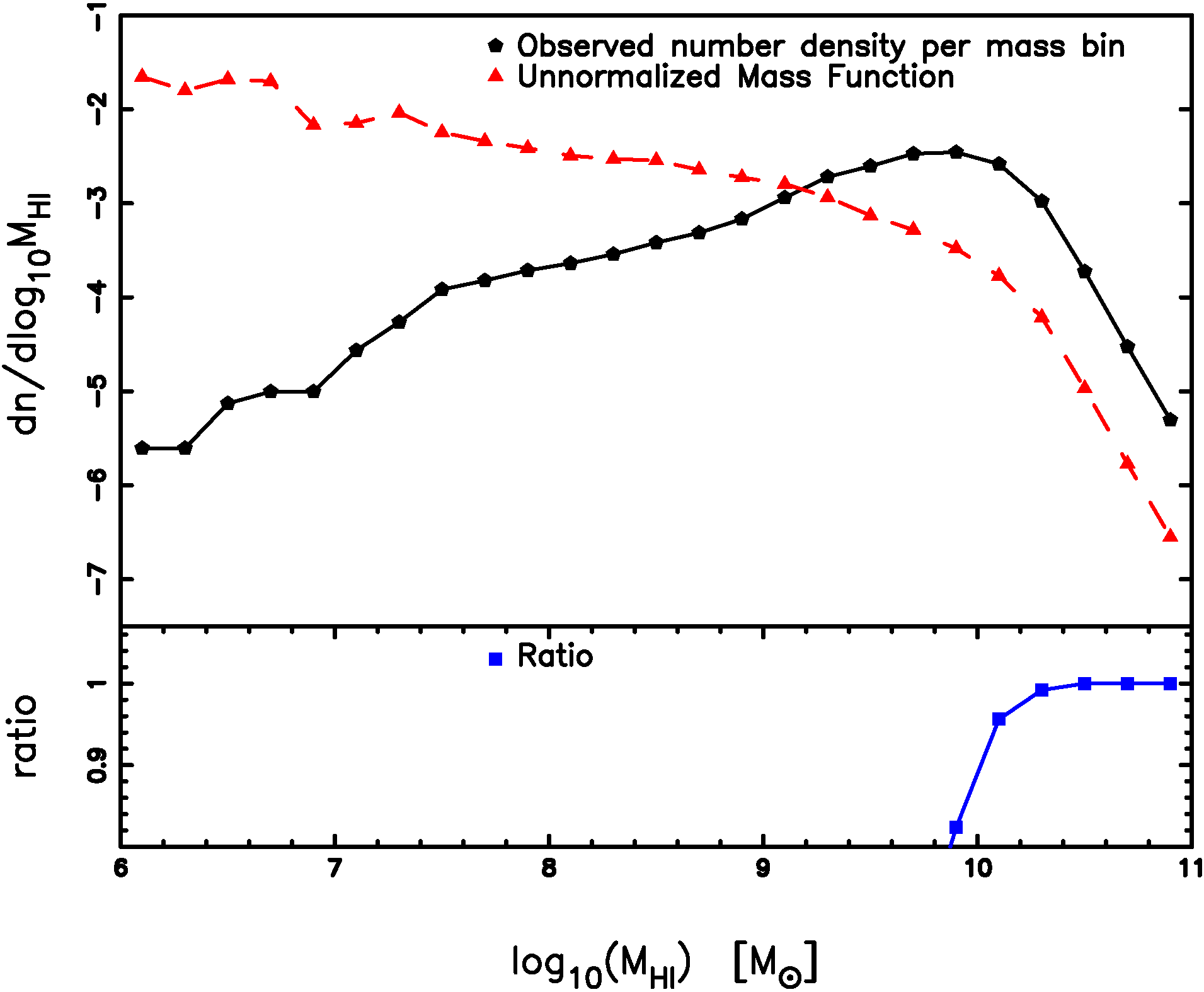}
\caption{Upper panel: Red dashed line is the mass function
estimated using 2DSWML method, which is not normalized. 
The black solid line is the number density per mass bin 
calculated for the same sample of HI galaxies.
Lower panel: The blue solid line shows the ratio
of the un-normalized mass function and the number density, 
multiplied by 100.}
\label{fig_ratio}
\end{figure}

One drawback for likelihood methods, as opposed to the $1/\vmax$ method,
is that the normalization of the HIMF is not fixed. 
This is obvious from eq.~\ref{eq_probi}.
There are a number of ways to fix the normalization 
\citep{1982ApJ...254..437D, 1997AJ....114..898W, 
  2003AJ....125.2842Z, 2010ApJ...723.1359M} which involve 
computing the selection function. Here we try a slightly 
different method. We assume that the high mass end of the 
HIMF is not affected by the selection function. One can test this assumption 
by comparing  the ratio of the normalized mass function from 2DSWML to that 
of the observed mass function which is related to the observed counts.
In the top panel of figure~\ref{fig_ratio} the observed HIMF is shown as filled pentagons
(solid line) and the un-normalized and converged HIMF from the 2DSWML is shown 
as filled triangles (dashed line). This is done for the $\alpha .100$ 
sample \citep{2018ApJ...861...49H}.
The ratio of these is shown for the last 7 mass bins
in the lower panel of figure~\ref{fig_ratio}. 
If the selection function affects the high mass end, the ratio
at this end would not have a flat feature. We indeed find that in this particular
example of $\alpha .100$ the last 3 mass bins are unaffected 
at the level ranging from $0-0.003\%$; whereas the 
last but third bin is relatively suppressed by around $0.4\%$. 
For this method to work we need to test the flatness of this ratio which means 
that at least the last two points at the high mass end should be unaffected 
by the selection function. Starting from the high mass end we progress sequentially 
to smaller bins which are unaffected by the 
selection function. The search is stopped when the 
selection function affects the particular bin at the level of $0.1\%$ or greater. Finally 
we fix the normalization by matching the integrated counts over these bins to that of the
observed HIMF.

\begin{figure}
\centering
\includegraphics[width=\columnwidth]{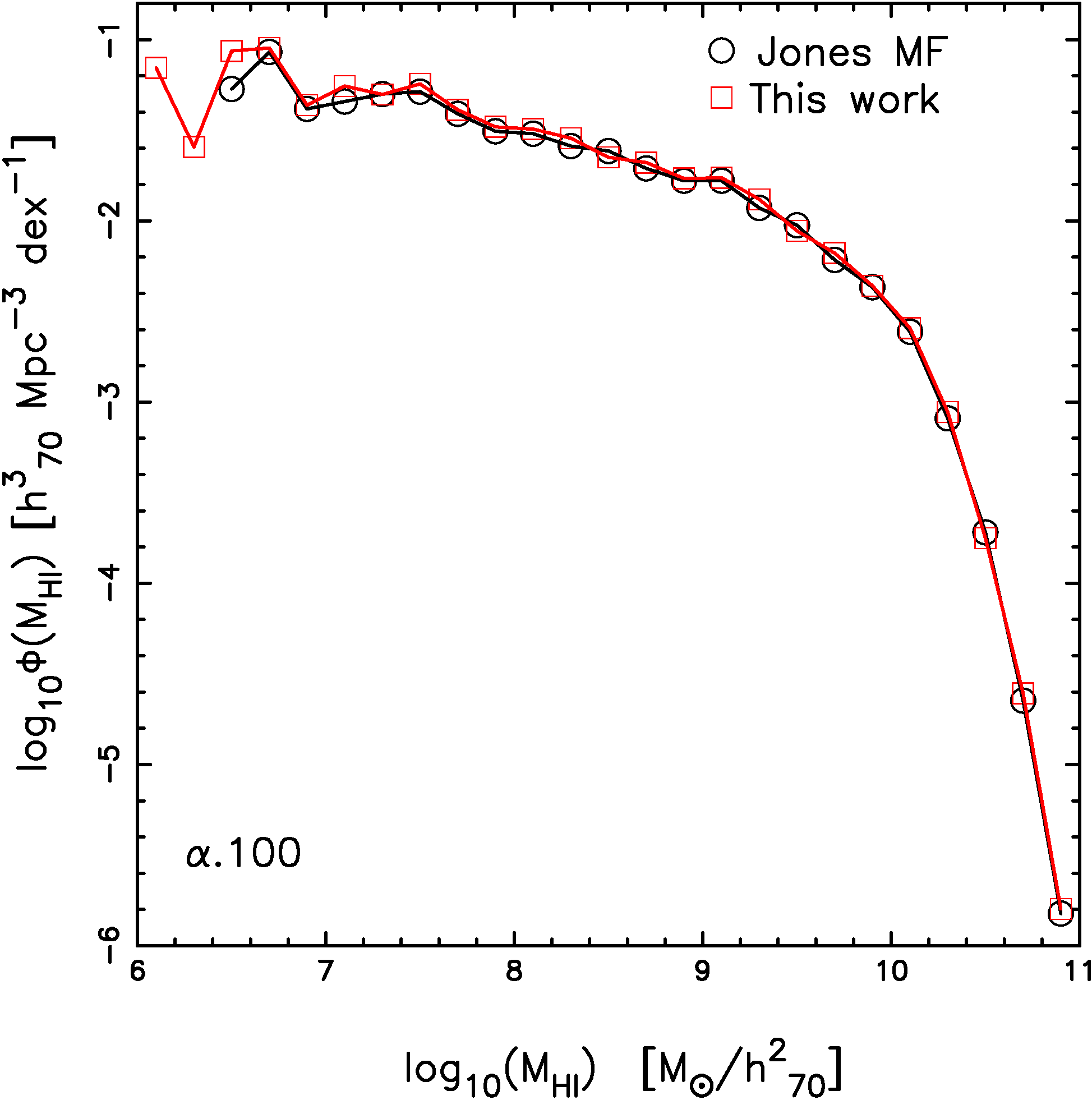}
\caption{Comparison of the HIMF in the $\alpha .100$ sample. The open circles 
is the HIMF by \citet{2018MNRAS.477....2J}. The open squares is the HIMF 
estimated by our implementation of the 2DSWML method.}
\label{fig_alfa100MF}
\end{figure}

We have compared this method to the one which normalizes the mass function 
to match the average observed counts as in 
\citet{1982ApJ...254..437D, 2010ApJ...723.1359M}. We find that
they match at the relative level of $\sim 0.4\%$. 
Finally we compare in figure~\ref{fig_alfa100MF} 
our result with that of $\alpha .100$ HIMF  \citep{2018MNRAS.477....2J}. As we can see
our implementation with some minor modifications reproduces the HIMF of 
\citep{2018ApJ...861...49H} extremely well.  

\subsection{Error Analysis of HIMF}

\quad \quad 
\emph{(i) Mass Errors: } Since $\mhi \propto S_{21} D^2$ the uncertainties on both 
integrated flux and distances lead to uncertainties on mass. 
Peculiar velocities of galaxies can affect distance estimates. This 
effect is larger in the local volume and smaller at higher redshifts where the Hubble
flow dominates over peculiar velocities. 
The $\alpha .40$ catalog also includes radial distances \citep{2011AJ....142..170H},  
using a local volume flow model \citep{2005PhDT.........2M} for galaxies 
with  $cz_{\text{\tiny{CMB}}} < 6000 \text{km.s}^{-1}$. This model has an estimated 
local velocity dispersion of $\sigma_{v}= 163 \text{km.s}^{-1}$. For these galaxies
we take the error in the distance to be the maximum  of $\sigma_v$ 
and 10$\%$ of the distance. For galaxies  $cz_{\text{\tiny{CMB}}} > 6000 \text{km.s}^{-1}$  
distances are estimated using the Hubble and we take the error on distances 
to be 10$\%$. Using the errors on distance and fluxes and their observed values 
we generate 300 Gaussian realizations and recompute $\mhi$ for every object.
We apply the 2DSWML method to all the realizations and find out the 
width of the distribution for every $\phi_j$ which we quote as an error. 

\emph{(ii)Poisson Errors:} 
The observed count in some of the bins is very low which means that 
it is important to consider errors due to Poisson counting.

\emph{(iii)Sample Variance:} We estimate this error by splitting
the area into 26 contiguous regions of  approximately equal area each. We compute 
the HIMF by removing one region at a time. Finally the  jackknife 
uncertainty for $\phi_j$ is computed as $\sigma_{\phi^j} = \frac{N-1}{N}\sum_{i=1}^{N=26} 
(\bar{{\phi}^j} - \phi^j_i)^2$ where $\bar{{\phi}^j}$ is the jackknife mean and 
 $\phi^j_i$ is the value for the i$^{\text{th}}$ jackknife sample.

\emph{(iv) Other Errors:} There are many other sources of errors which 
are discussed in \citet{2018MNRAS.477....2J}. E.g. the error associated 
with the 2DSWML method which one can either estimate using the information matrix
\citep{1988MNRAS.232..431E} or by making further mocks \citep{2018MNRAS.477....2J}.
One needs to understand how these errors are correlated with a finite sample 
or Poisson errors. We also expect Poisson errors to be correlated to mass errors 
especially in the lowest and highest mass bins where the observed counts are low.
Another source of uncertainty is the error in the velocity width $\w50$.
Since the peak flux, $S_{\text{peak}}$, is inversely proportional to $\w50$ 
we expect their errors to be correlated. Since we integrate over $\w50$ 
to obtain the HIMF, we do not consider errors in $\w50$.  
In order to properly account for errors
one will need their covariances. We have attempted to add further sources of errors 
in quadrature but we find that the error bars become progressively larger
and the Schechter function fits have a $\chi^2_{\text{red}} < 1$ which means that 
we may be overestimating the errors. For this work we stick to the errors $(i)-(iii)$
and add them in quadrature. These errors are comparable to 
\citet{2010ApJ...723.1359M, 2011AJ....142..170H}.

\label{lastpage}

\end{document}